\RequirePackage{fix-cm}
\documentclass[twocolumn,epjc3]{svjour3}  
\smartqed  
\RequirePackage{graphicx}
\journalname{Eur. Phys. J. C}
\usepackage[usenames]{color}
\usepackage{colortbl}
\usepackage{multirow}
\begin{document}

\title{Observation of the temperature and barometric effects on the cosmic muon flux by the DANSS detector
}


\author{I.~Alekseev\thanksref{addr1, addr2, addr3}
        \and
        V.~Belov\thanksref{addr4}
        \and
        V.~Brudanin\thanksref{addr4, d1}
        \and
        A. Bystryakov\thanksref{addr4, addr6}
        \and
        M.~Danilov\thanksref{addr5}
        \and
        V.~Egorov\thanksref{addr4, addr6, d1}
        \and
        D.~Filosofov\thanksref{addr4}
        \and
        M.~Fomina\thanksref{addr4}
        \and
        S.~Kazartsev\thanksref{addr4, addr9}
        \and
        A.~Kobyakin\thanksref{addr1, addr3}
        \and
        A.~Kuznetsov\thanksref{addr4}
        \and
        I.~Machikhiliyan\thanksref{addr10}
        \and
        D.~Medvedev\thanksref{addr4}
       	\and
       	V.~Nesterov\thanksref{addr1}
       	\and
       	I.~Rozova\thanksref{addr4}
       	\and
       	N.~Rumyantseva\thanksref{addr4}
       	\and
       	V.~Rusinov\thanksref{addr1}
       	\and
       	E.~Samigullin\thanksref{addr1, e1}
       	\and
       	Ye.~Shevchik\thanksref{addr4}
       	\and
       	M.~Shirchenko\thanksref{addr4}
       	\and
       	Yu.~Shitov\thanksref{addr4}
       	\and
       	N.~Skrobova\thanksref{addr1, addr5}
       	\and
       	A.~Starostin\thanksref{addr1}
       	\and
       	D.~Svirida\thanksref{addr1, addr3, addr5}
       	\and
       	E.~Tarkovsky\thanksref{addr1}
       	\and
        E.~Yakushev\thanksref{addr4}
	\and
	I.~Zhitnikov\thanksref{addr4}
	\and
	D.~Zinatulina\thanksref{addr4}
}

\thankstext{e1}{e-mail: eduk007@yandex.ru}
\thankstext[$\star$]{d1}{Deceased}


\institute{Alikhanov Institute for Theoretical and Experimental Physic NRC "Kurchatov Institute", B. Cheremushkinskaya str. 25, Moscow, 117218, Russia \label{addr1}
           \and
National Research Nuclear University MEPhI (Moscow Engineering Physics Institute), Kashirskoe highway 31, Moscow, 115409, Russia \label{addr2}
           \and
Moscow Institute of Physics and Technology, Institutskiy lane 9, Dolgoprudny, Moscow Region, 141701, Russia \label{addr3}
			\and
Joint Institute for Nuclear Research, Joliot-Curie str. 6, Dubna, Moscow region, 141980, Russia \label{addr4}
            \and
Dubna State University, Universitetskaya str. 19, Dubna, Moscow Region, 141982, Russia \label{addr6}
			\and
Lebedev Physical Institute of the Russian Academy of Sciences, Leninskiy avenue 53, Moscow, 119991, Russia \label{addr5}
			\and
Voronezh State University, Universitetskaya square 1, Voronezh, 1394018, Russia \label{addr9}
            \and
Federal State Unitary Enterprise Dukhov Automatics Research Institute, Sushchevskaya 22, Moscow 127055, Russia \label{addr10}
}

\date{Received: date / Accepted: date}

\maketitle

\begin{abstract}
The DANSS detector~\cite{danss} is located directly below a commercial reactor core at the Kalinin Nuclear Power Plant. Such a position provides an overburden about 50 m.w.e. in vertical direction. In terms of the cosmic rays it occupies an intermediate position between surface and underground detectors. The sensitive volume of the detector is a cubic meter of plastic scintillator with fine segmentation and combined PMT and SiPM readout, surrounded by multilayer passive and active shielding. The detector can reconstruct muon tracks passing through its sensitive volume. The main physics goal of the DANSS experiment implies the antineutrino spectra measurements at various distances from the source. This is achieved by means of a lifting platform so that the data is taken in three positions - 10.9, 11.9 and 12.9 meters from the reactor core. The muon data were collected for nearly four calendar years. The overburden parameters $\langle E_{thr}\cos\theta \rangle$ and $\langle E_{thr} \rangle$, as well as the temperature and barometric correlation coefficients are evaluated separately for the three detector positions and, in each position, in three ranges of the zenith angle - for nearly vertical muons with $\cos\theta>0.9$, for nearly horizontal muons with $\cos\theta<0.36$, and for the whole upper hemisphere.
\end{abstract}

\section{Introduction}
\label{intro}  

Cosmic radiation, and particularly the muon flux at the ground level, is an important instrument for the Universe investigations. At the same time, cosmic muons and muon-induced processes are the main background for neutrino experiments, especially for those located on the surface level or at shallow depths. Some even deep-underground detectors, aimed, for example, at the dark matter searches, are looking for tiny effects with daily or seasonal periodicity. All these experiments need reliable predictions of the muon flux for background estimates or weather effect compensations. The models should be versatile, work at different depths or altitudes, cover all angular regions of the sky and accurately account for meteorological changes in the atmosphere. Many experiments capable of muon detection and identification provide data for various checks of existing theoretical models. Yet the DANSS experiment with its specific position in terms of the overburden parameter and remarkable tracking capability can make an important contribution to the further development of the theoretical approaches.

In the middle of the 20th century scientists discovered the dependence between the flux of the penetrating component of the cosmic rays and some of the atmospheric parameters. Now it is known that this penetrating component mainly consists of the muons which are produced in decays of charged mesons, and the most significant meteorological influence comes from the temperature and the barometric effects. The barometric effect consists of three main terms \cite{sagisaka1986}. First of all with higher pressure the air pillar above the observation point is also higher, so it increases the altitude of the muon production, and they have to fly through more air, so that the muon absorption grows. Also the muons have to travel a greater distance to reach the ground, which leads to the growth of $\mu$-e decays. From the other hand larger amount of air increases the pion and, accordingly, the muon production. The last reason prevails at high altitude, and becomes negligibly small on the sea level. The temperature effect has two opposite terms \cite{barret}: positive and negative. With the growth of the temperature the atmospheric density decreases, so the interaction probability of the mesons also decreases and thus larger fraction of them decays to muons. At the same time, as the atmosphere expands, the muons are produced at higher altitude, and their low energy fraction has higher probability to decay before reaching the ground. The numeric values characterizing these effects are usually denoted as $\alpha$ and $\beta$ – the correlation coefficients between the relative variation of the muon flux and that of the effective temperature of the atmosphere, or of the pressure
on the ground level, in absolute terms, accordingly. Both coefficients depend on the overburden, and there are many experiments either far underground
or on the ground surface which measured the corresponding values. The DANSS experiment, though in fact elevated 7 m above the surface, is, at the same time, only a few meters below a mighty reactor core with all its biological shielding, and can access $\alpha$ and $\beta$ in the intermediate area of the overburden parameter, corresponding to very shallow depth. With its good tracking capability, fairly good angular resolution and huge statistics, DANSS can make accurate measurements in several well defined regions of the zenith angle.

\section{The DANSS detector}
\label{sec:1}

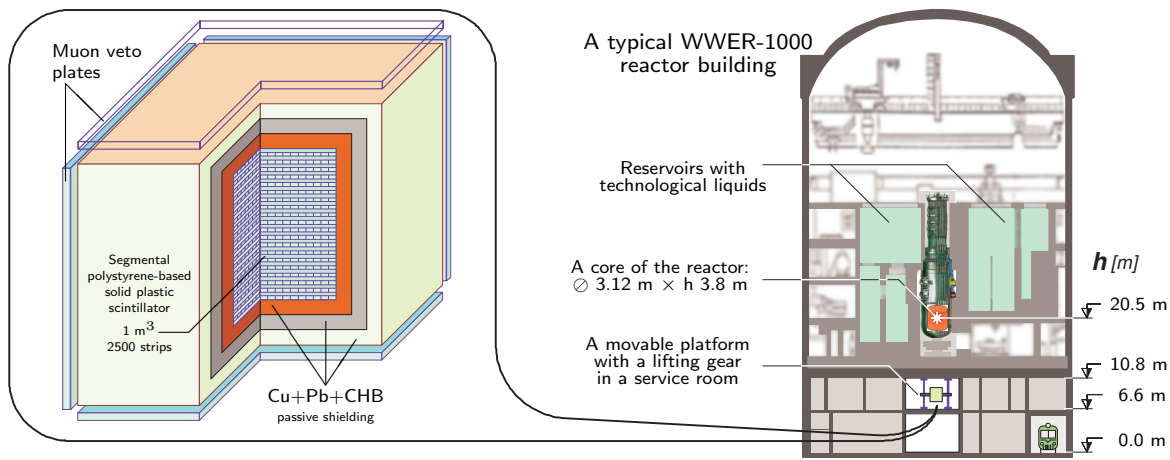
\begin{figure*}
	\setlength{\unitlength}{1mm}
	\begin{center}
	\begin{picture}(150,60.0)(0,0)
		\put(8,5){\includegraphics[scale=0.9]{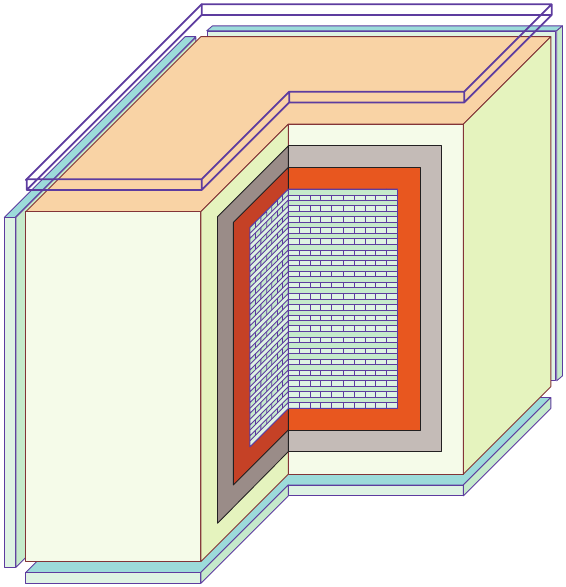}}
		\put(11,20){\parbox{15mm}{\begin{center}\tiny\sf Segmental polystyrene-based solid plastic scintillator\\[1mm]
					1~m$^3$\\2500 strips \end{center}}}
		\put(21,17){\line(1,0){4}}
		\put(25,17){\line(1,1){10}}
		\put(43,9.5){\makebox(0,0)[t]{\scriptsize\sf Cu+Pb+CHB}}
		\put(43,6.5){\makebox(0,0)[t]{\tiny\sf passive shielding}}
		\put(43,10){\line(-2,3){7}}
		\put(43,10){\line(0,1){8.5}}
		\put(43,10){\line(2,3){4}}
		\put(7,52){\parbox{15mm}{\scriptsize\sf Muon veto\\plates}}
		\put(8.75,50){\line(0,-1){13}}
		\put(8.75,50){\line(1,-1){5}}
		\put(100.0,0){\includegraphics{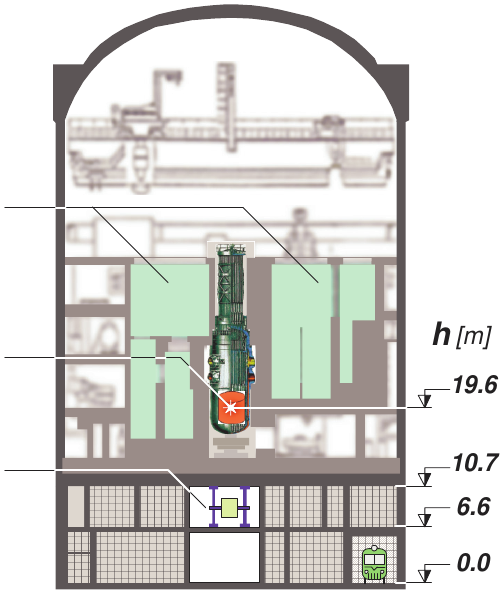}}
		\put(0.9,2.2){\includegraphics{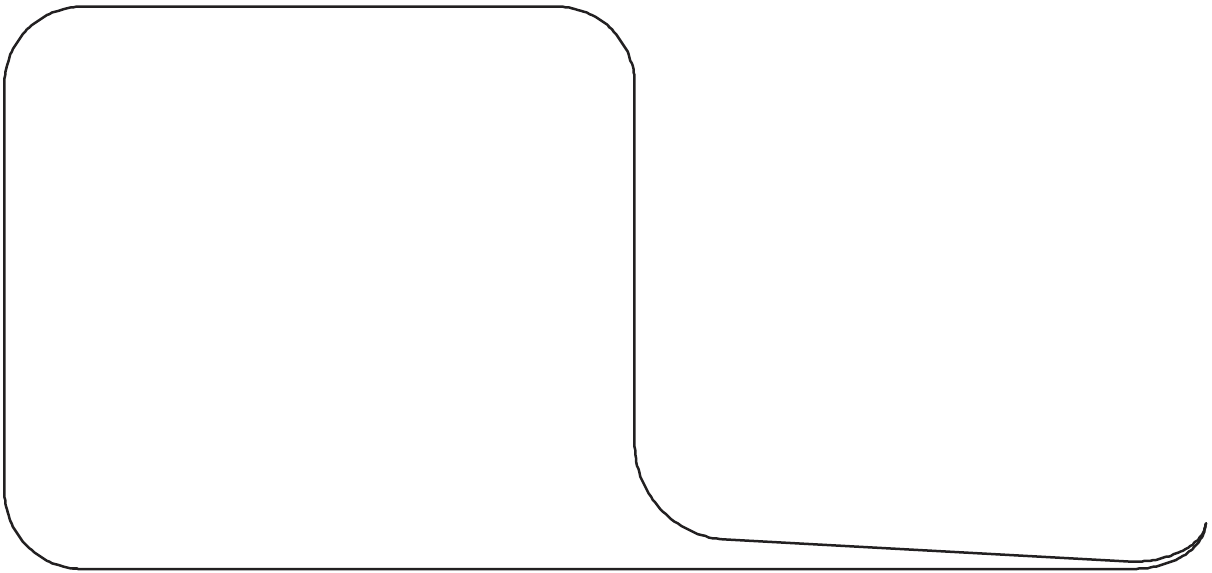}}
		\put(72,53.0){\parbox{40mm}{\begin{center}\footnotesize\sf A typical WWER-1000 \\[-0.5mm]reactor building\end{center}}}
		\put(75,36.7){\parbox{30mm}{\begin{center}\scriptsize\sf Reservoirs with\\[-0.5mm] technological liquids\end{center}}}
		\put(72,23.7){\parbox{30mm}{\begin{center}\scriptsize\sf A core of the reactor:\\[-0.5mm] $\oslash$ 3.12 m $\times$ h 3.8 m\end{center}}}
		\put(73,12.5){\parbox{30mm}{\begin{center}\scriptsize\sf A movable platform\\[-0.5mm] with a lifting gear\\[-0.5mm] in a service room\end{center}}}
		\put(135,20){\parbox{30mm}{\begin{center}\scriptsize\sf \fcolorbox{white}{white}{20.5 m}\end{center}}}
		\put(135,12){\parbox{30mm}{\begin{center}\scriptsize\sf \fcolorbox{white}{white}{10.8 m}\end{center}}}
		\put(135,8){\parbox{30mm}{\begin{center}\scriptsize\sf \fcolorbox{white}{white}{ 6.6 m}\end{center}}}
		\put(135,2){\parbox{30mm}{\begin{center}\scriptsize\sf \fcolorbox{white}{white}{ 0.0 m}\end{center}}}
	\end{picture}
	\caption{Layout of the DANSS detector and its location in the reactor building.}
	\label{fig:1}
	\end{center}
\end{figure*}

 DANSS (Detector of AntiNeutrino based on Solid Scintillator) is designed for the searches of the short-range neutrino oscillations. It is placed on a lifting system under an industrial 3.1 GW$_{th}$ reactor of the Kalinin Nuclear Power Plant (57.91$^o$ N, 35.06$^o$ E). The layout of the DANSS detector and the illustration of its installation are shown in Figure~\ref{fig:1}. The reactor building above the detector provides $\sim$ 50 m.w.e. of passive shielding against the cosmic radiation in vertical direction. The data is typically taken in 3 positions of the lifting system: up, middle and down – 10.9 m, 11.9 m, 12.9 m from the reactor core center to the detector center, respectively. DANSS is a highly segmented plastic detector with 1 m$^3$ of total sensitive volume, and consists of 100 layers of 25 scintillator strips each. The strips in the adjacent layers lay in perpendicular directions. Every strip (1 $\times$ 4 $\times$ 100 cm$^3$) is made of scintillating plastic with thin gadolinium-loaded coating for neutron capture and has 3 wavelength-shifting fibers glued into grooves along the strip. Each strip is equipped with an individual silicon photomultiplier (SiPM), and every 50 parallel strips are grouped into modules which have one common photomultiplier tube (PMT) for the light collection. The sensitive volume is surrounded by several layers of passive and active shielding: copper (5 cm), borated polyethylene (8 cm), lead (5 cm), borated polyethylene (8 cm), and a muon scintillation veto system on 5 sides. The details on the detector design and on the background conditions in the detector hall can be found in~\cite{danss}.

\section{Muon data}

\subsection{Muon selection}

\begin{figure*}
	\begin{minipage}[h]{0.49\linewidth}
		\center{\includegraphics[width=1\linewidth]{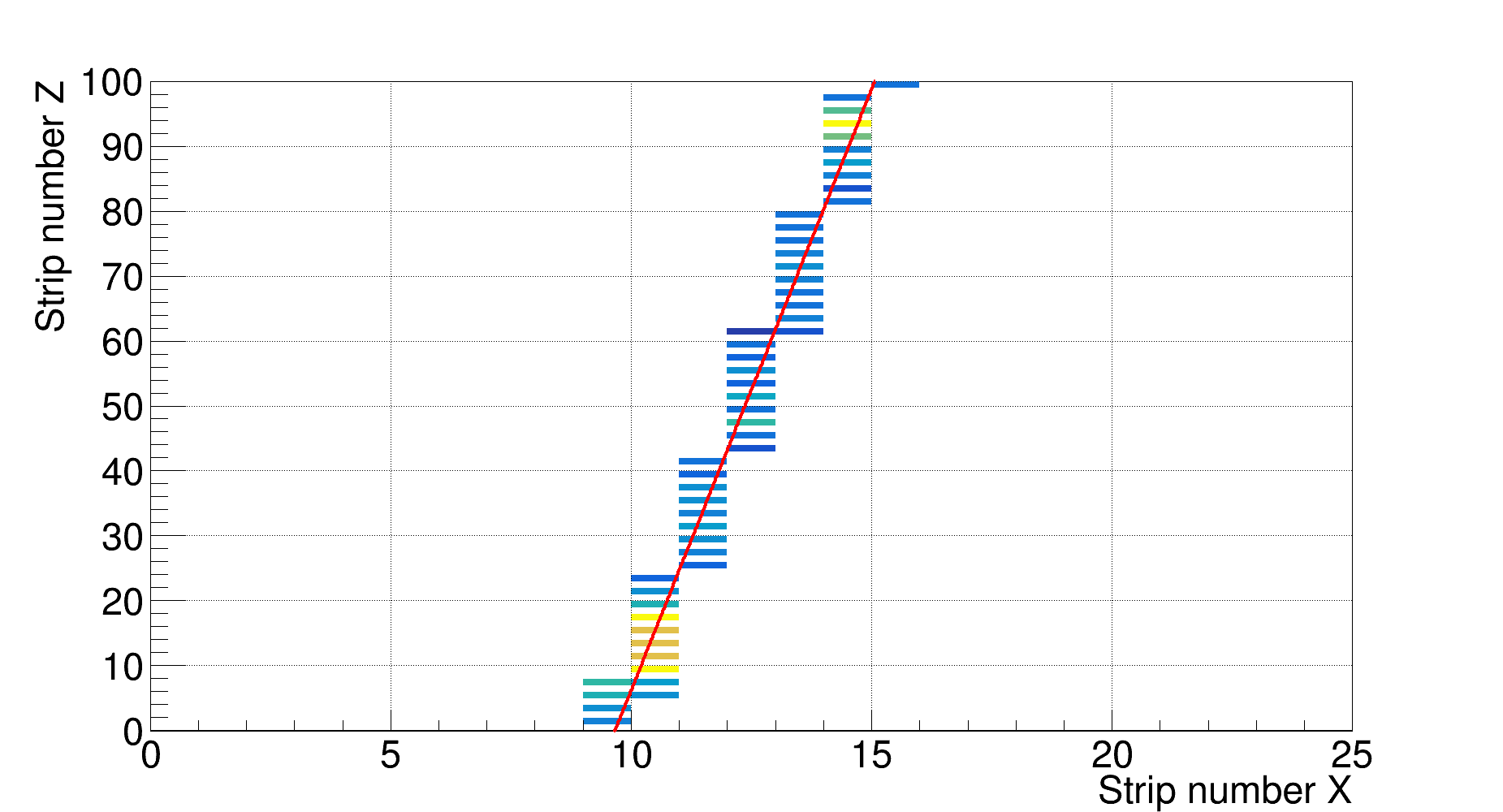} \\ a)}
	\end{minipage}
	\hfill
	\begin{minipage}[h]{0.49\linewidth}
		\center{\includegraphics[width=1\linewidth]{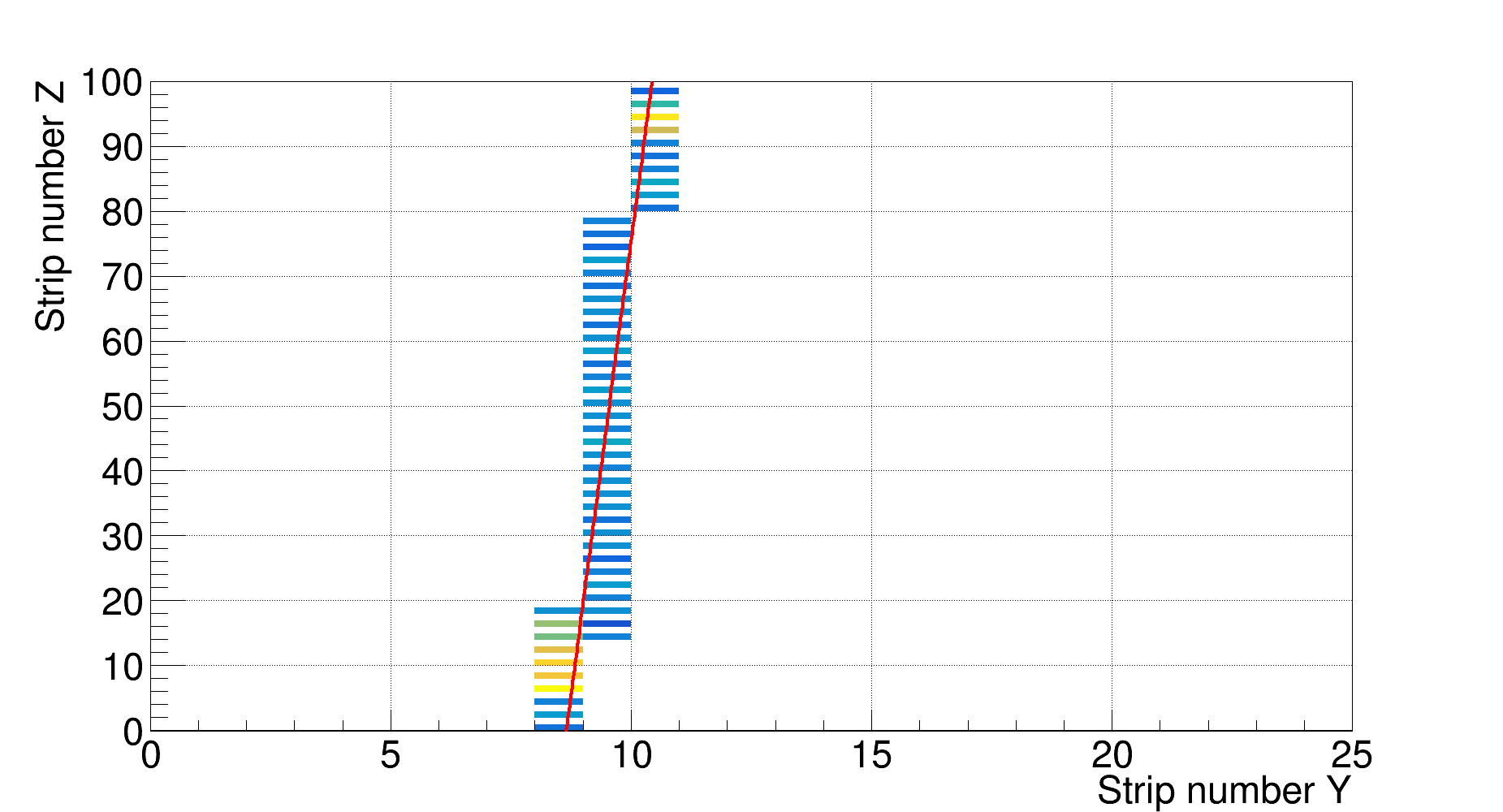} \\ b)}
	\end{minipage}
	\caption{Hits on a muon track as seen from the two sides of the detector. The colored rectangles indicate the triggered strips and the red lines show the reconstructed muon track.}
	\label{fig:2}
\end{figure*}

\begin{figure}
  \includegraphics[width=0.49\textwidth]{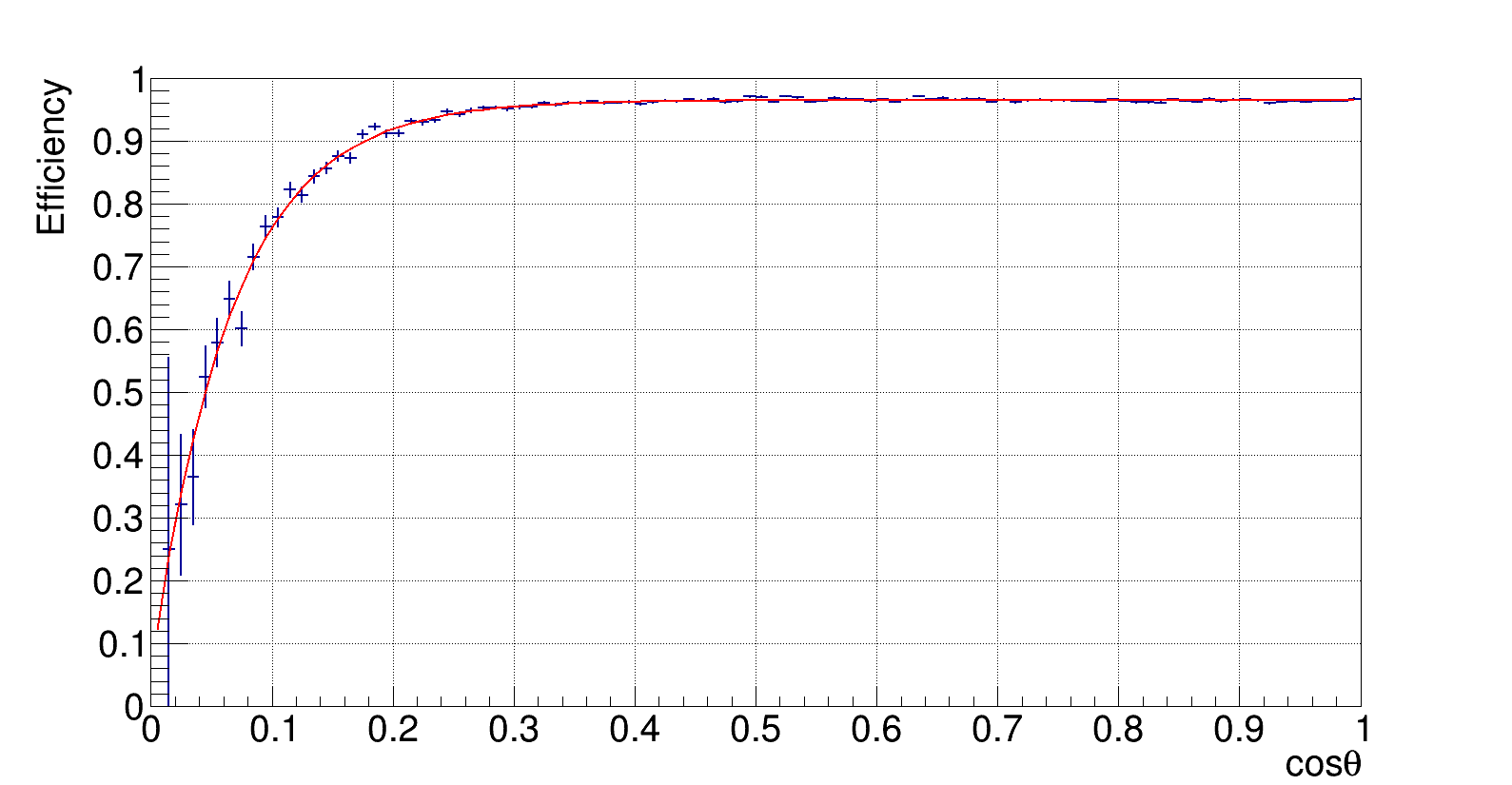}
\caption{Dependence of the muon reconstruction efficiency on the cosine of the zenith angle $\theta$ and its polynomial representation (red).}
\label{fig:3}
\end{figure}

This research covers the period of almost continuous data taking from October 5, 2016 till September 1, 2020. 
The hadronic and light electromagnetic components of the cosmic rays are not strongly pronounced already on the surface level (see sect. 30.3 of~\cite{pdg} for example). Overburden of 50 m.w.e. arising from the reactor building and its infrastructure, adds orders of magnitude to the suppression of these components, compared to the muon flux. Muon-induced background at this depth mainly consists of neutrons~\cite{hausser}, which are not numerous and have quite different signature in the detector. As a result any straight track above the minimal length can be associated with a cosmic muon. Effective reconstruction of such tracks becomes possible due to the high segmentation of the detector and to the individual SiPM readout. The cubic shape of the detector is not very convenient for the computation of solid angles and corresponding muon fluxes. To avoid this complexity only muons that fly closer than 40 cm from the detector center are used in the further analysis, at the same time providing long enough tracks for precise calculations of the track parameters. A dedicated algorithm for the searching of such kind of tracks and for the calculation of their parameters was developed. In Figure~\ref{fig:2} one can see an example of hits on a muon track and a reconstructed straight line. The performance of this algorithm was verified with a Monte-Carlo simulation and the behavior of the reconstruction efficiency is shown in Figure~\ref{fig:3}. Due to the discreteness of the detector, and to the fact that the sensitive strips have horizontal direction, the registration efficiency decreases for the nearly horizontal muons, given the algorithm requires some minimal number of strip layers to be hit by a muon track. However the total efficiency is about 97\%, if averaged with the angular dependence of the muon flux.  A fit of this dependence, red in Figure~\ref{fig:3}, is used in the further calculations.

\begin{figure}
  \includegraphics[width=0.49\textwidth]{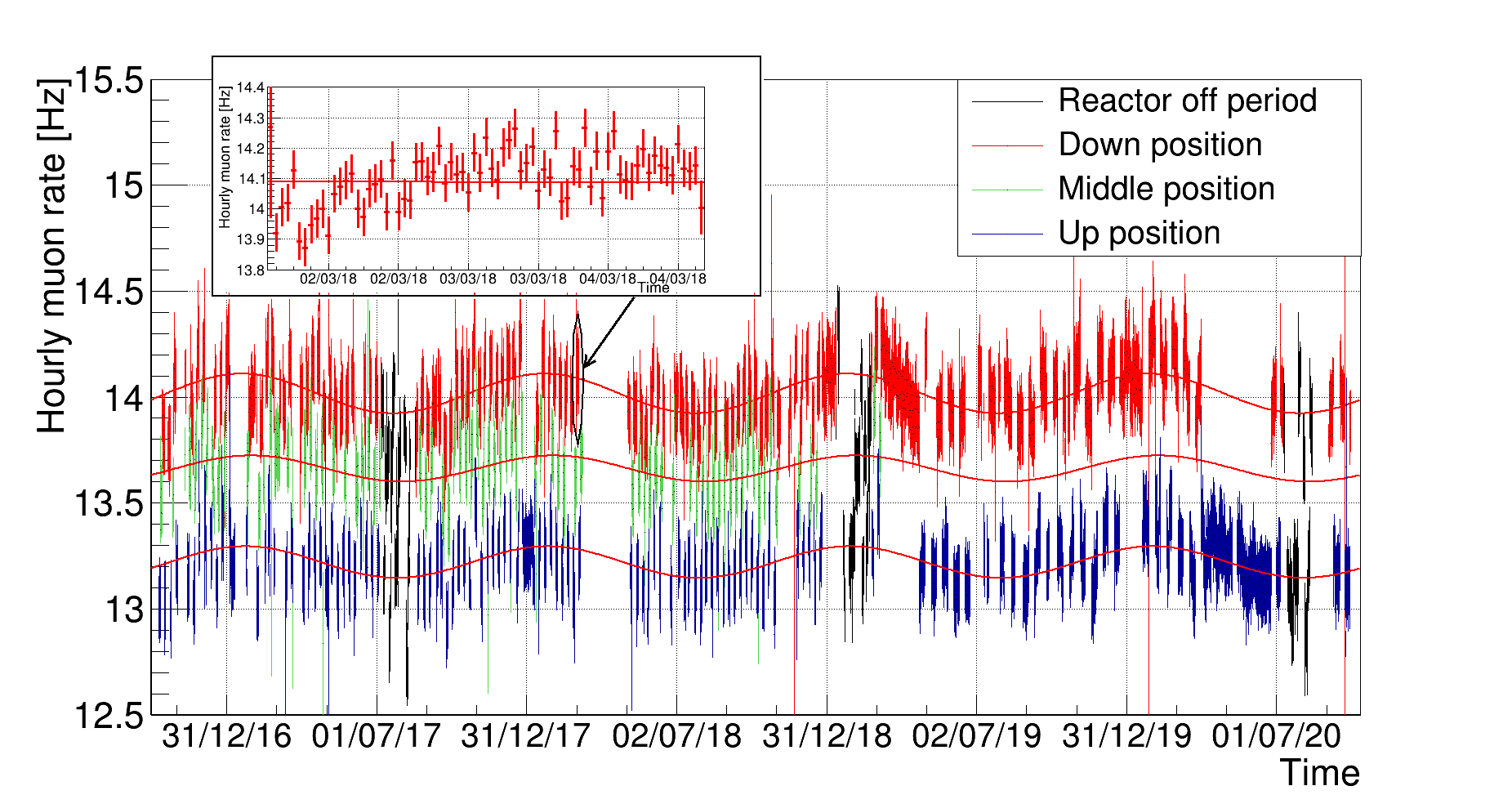}
	\caption{Hourly muon flux over the calendar time for different detector positions: blue, green and red are for up, middle and down positions, black – excluded data of the reactor-off periods. Solid red lines show sinusoidal fits of these distributions. The insertion shows the fine structure of a three day interval with the detector in the down position.}
	\label{fig:4}
\end{figure}

Hourly muon fluxes calculated using the above algorithm for each of the detector positions are presented in Figure~\ref{fig:4}. Only a small seasonal variation is observed, and this is strongly blurred by the barometric effect. The latter takes place because the averaged ground pressure typically does not depend on the season (see Figure~\ref{fig:10} below), while its changes on a time scale of hours or days are significant. The data were fitted by sinusoidal functions with the period fixed to one year. The seasonal variation is less than 1$\%$ of the average flux, and the yearly minimum falls approximately on the last day of July. Black areas on the plot correspond to the periods of the reactor fuel reloads when about 10 meters of water were added to the pools above the detector. Data from these periods were excluded from the further analysis. The insertion with the zoomed view in Figure~\ref{fig:4} gives an illustration of the statistical errors of hourly flux measurements.
\subsection{Angular distributions of the muon flux}
\begin{figure}
	\includegraphics[width=0.49\textwidth]{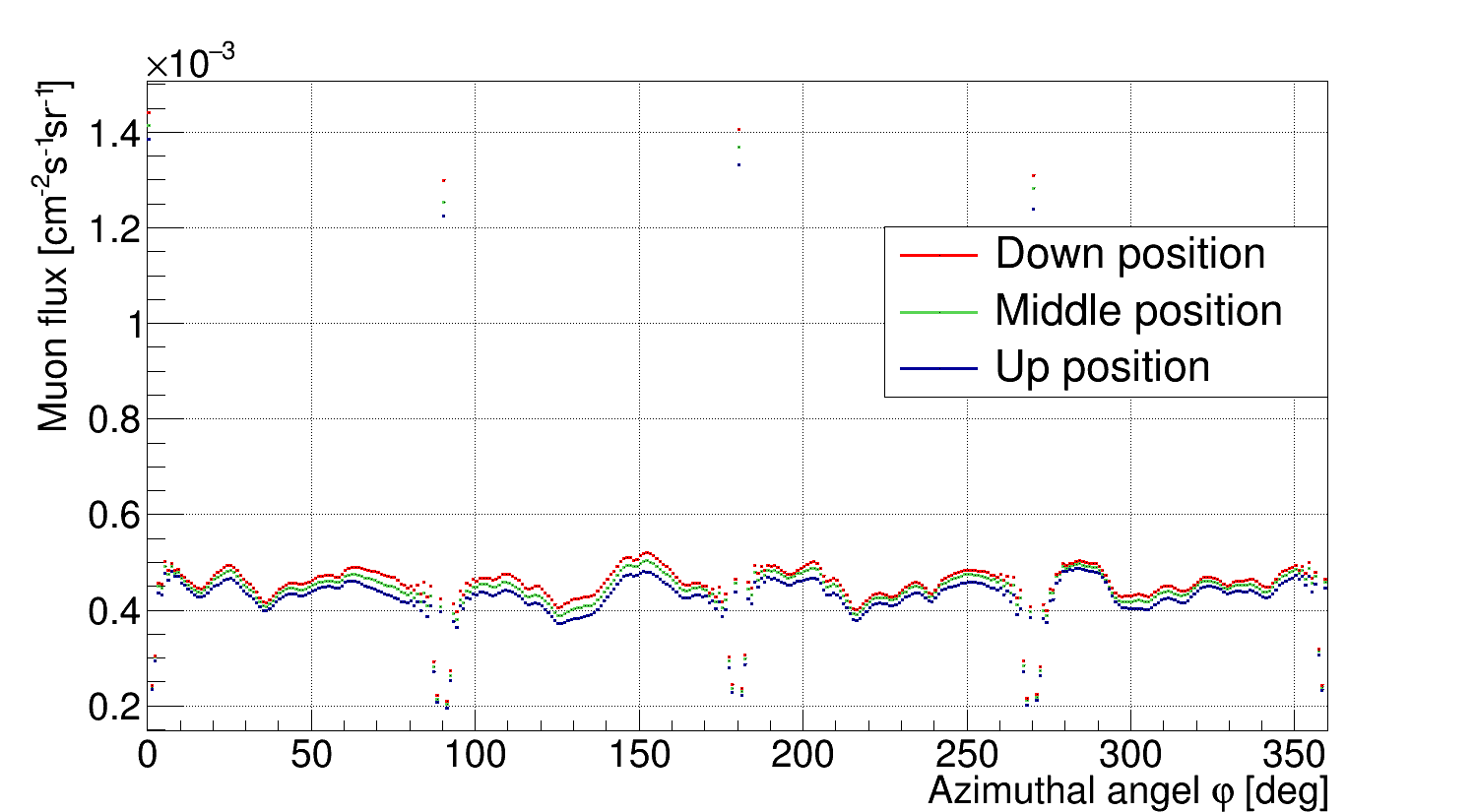}
	\caption{Dependencies of the observed muon flux on the azimuthal angle in each detector position: blue, green and red are for up, middle and down positions.}
	\label{fig:5}
\end{figure}

Based on the tracking algorithm the azimuthal distributions of the muon flux were built for each of the detectors positions, and they are shown in Figure~\ref{fig:5}. Significant jumps at the azimuthal angles $\phi$ = 0$^o$, 90$^o$, 180$^o$ and 270$^o$ and corresponding dips around these values could be explained by the fact that the strips lay along these directions. A muon can fly diagonally through a single vertical stack of strips with the displacement of the azimuthal angle $|$$\Delta$$\phi$$|$\textless arctan(0.04)$\approx$2.3$^o$, in which case it is indistinguishable from a muon going at corresponding angles of exact multiple of 90$^o$. All other angular intervals show only small deviations from a constant value, thus reflecting the anisotropy of the reactor infrastructure above the detector. All in all, the distributions are quite flat, and in the further analysis they will be considered as such.

\begin{figure}
\includegraphics[width=0.49\textwidth]{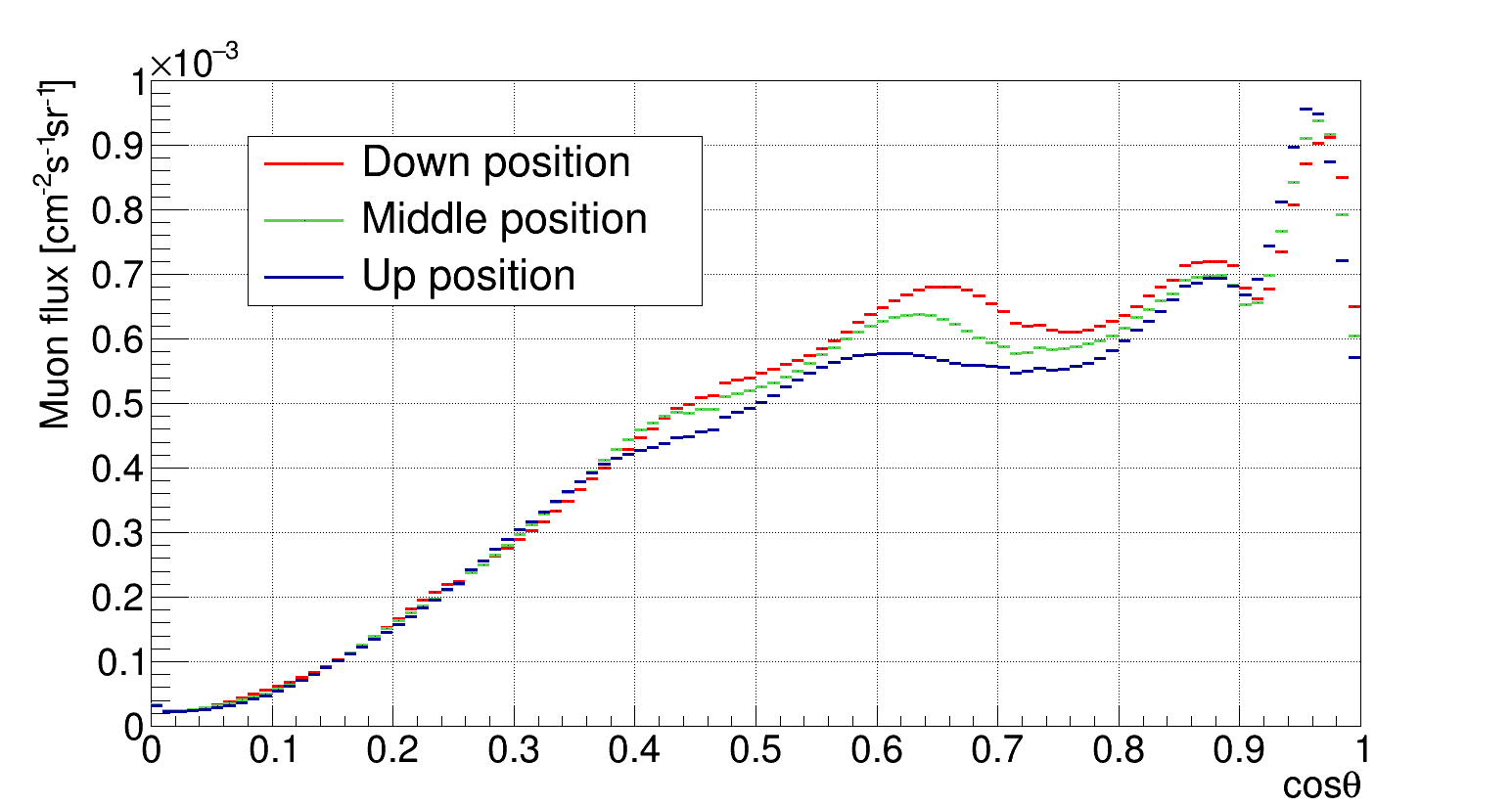}
\caption{Dependence of the experimental muon flux on the cosine of the zenith angle in each of the detector positions: blue, green and red are for up, middle and down positions.}
\label{fig:6}
\end{figure}

The measured dependence of the muon flux on the cosine of the zenith angle is presented in Figure~\ref{fig:6} for each of the three detector positions. The shadowing of the cosmic muons mainly happens from the water pools near the reactor, the reactor body and the surrounding shielding, which is located straight above the detector. When the detector distance from the core grows, the shape of the shadow changes accordingly, together with significant changes to the angular distribution. This fact is taken into account in the further analysis: the estimates are made separately for each detector position. The peak of the flux for the nearly vertical muons is associated with the gap between the reactor body and the walls of the biological shielding of the reactor.

\subsection{Muon threshold energy}

The correlation coefficients depend on the overburden, and one of the values which characterizes this fact is the muon threshold energy $E_{thr}$. This is the minimum energy that a muon must have on the surface level to fly through all the matter between the surface and the detection point. The measurement of $E_{thr}$ is very important because it is needed for the calculation of the effective temperature and should be compared to the results of other experiments and theory predictions. The design of the reactor building is too complicated to perform a precise Monte-Carlo simulation of the muons passing through the building structure. The method proposed in this paper is based on the comparison of the phenomenological predictions to the experimental distributions of the muon flux at different values of the zenith angle. The predictions are based on the differential cosmic muons spectra at the surface level and two independent variants of these spectra were chosen for the comparison. The first one is taken from \cite{spec1}:
\begin{eqnarray}
	\label{eq:1} 
	\frac{dI\left(\theta, \phi, p \right)}{d\theta d\phi dp}=\hspace{16em}\nonumber \\
	=\frac{18\left(p+5\right)}{\left(p\cos\theta+145\right)\left(p+5\sec\theta\right)\left(p+2.7\sec\theta\right)^{2.7}},
\end{eqnarray}
where p is the muon momentum measured in GeV/c in the interval 1$\leq$p$\leq$10$^5$ GeV. The second variant comes from \cite{gaisser}: 
\begin{eqnarray}
	\frac{dN_\mu}{dE_\mu}=\frac{0.14E_\mu^{-2.7}S\left(E_\mu\right)}{cm^2\times s\times sr\times GeV}\times\nonumber\\
	\times\left\{	\frac{1}{1+\frac{1.11E_\mu \cos\theta}{115 GeV}}+\frac{0.054}{1+\frac{1.11E_\mu \cos\theta}{850 GeV}}\right\},
\end{eqnarray}
where $S(E_\mu$) is the suppression factor which shows changes in the muon flux because of the muon decays and their energy loss, and at the sea level it could be written as:

\begin{eqnarray}
	\label{eq:3} 
	S\left(E_{\mu}\right)=\left(\frac{\Lambda_N\cos\theta}{X_0}\right)^{p_1}\times\hspace{8em}\nonumber\\
	\times\left(\frac{E_\mu}{E_\mu+2GeV/\cos\theta}\right)^{p_1+\gamma+1}\Gamma\left(p_1+1\right)
\end{eqnarray}
where, in turn, $p_1=\epsilon_\mu⁄\left(E_\mu \cos\theta+\kappa X_0\right)$, $\epsilon_\mu$=1 GeV is the muon critical energy, $X_0$ is the observation depth, for the sea level $X_0\approx$ 1030 g/cm$^2$, $\kappa$=2 MeV/g/cm$^2$ is the muon energy loss as of a minimum ionizing particle, $\Lambda_N$ – attenuation length of the primary cosmic rays and $\gamma$ is the muon spectral index, refer to the Table~\ref{tab:2} for the omitted numerical values.

\begin{figure*}
	\begin{minipage}[h]{0.49\linewidth}
		\center{\includegraphics[width=1\linewidth]{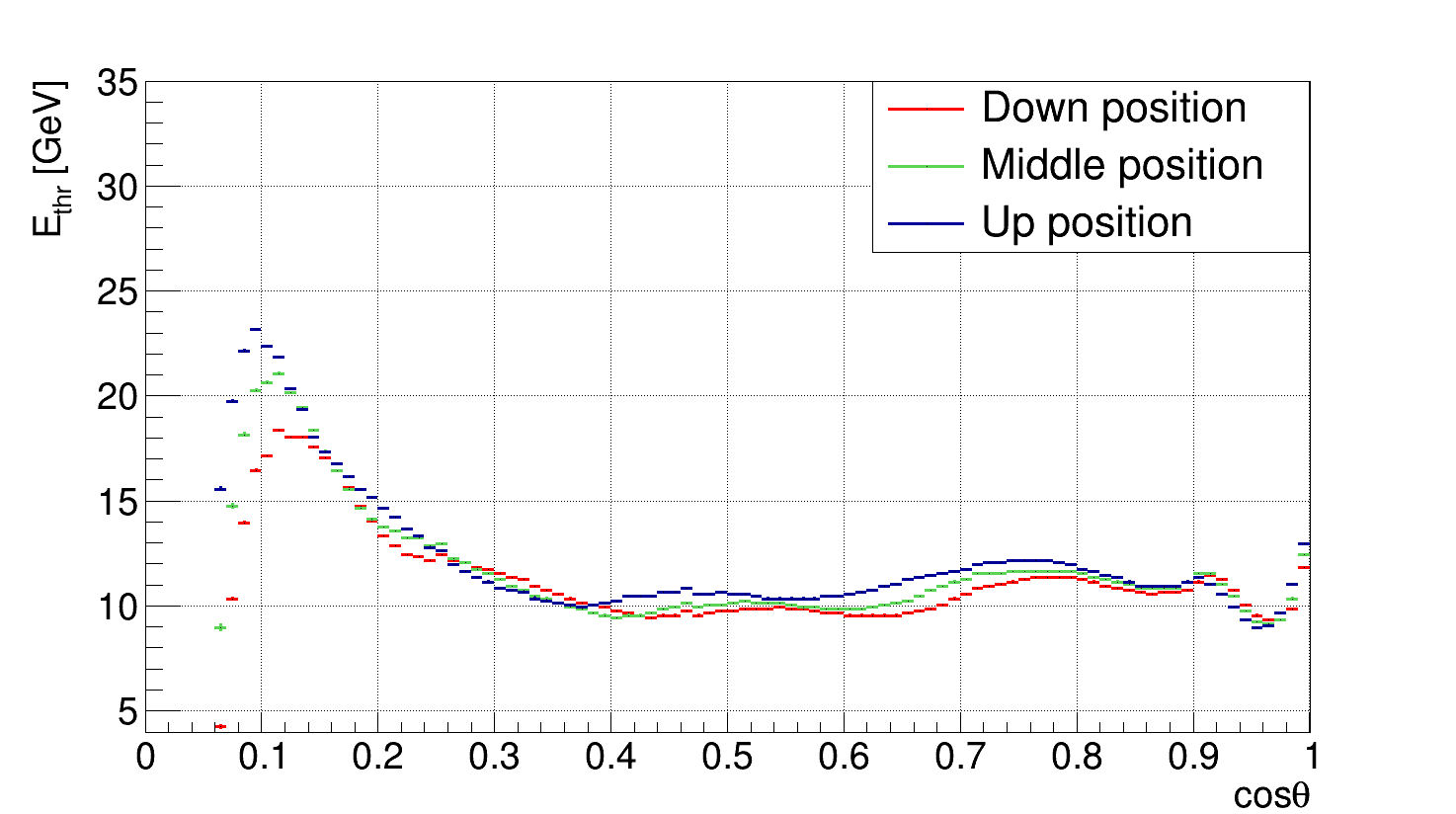} \\ a)}
	\end{minipage}
	\hfill
	\begin{minipage}[h]{0.49\linewidth}
		\center{\includegraphics[width=1\linewidth]{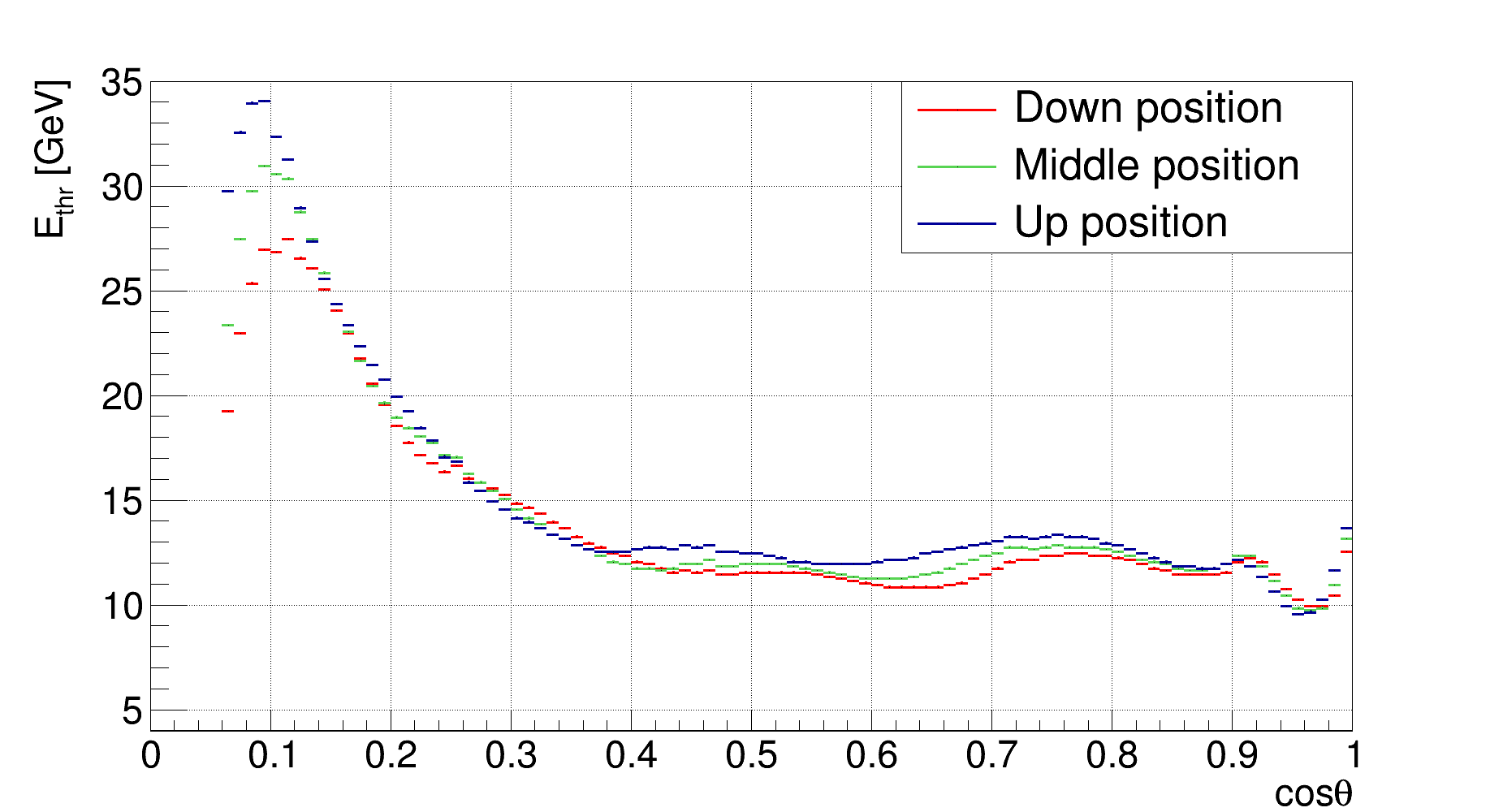} \\ b)}
	\end{minipage}
	\caption{Calculated $E_{thr}$ for every $\cos\theta$ based on the spectrum \cite{spec1} (a) and on the spectrum \cite{gaisser} (b), for each detectors position: blue, green and red are for up, middle and down positions. The area with $\cos\theta < 0.06$ is excluded from the plot because of the low detection efficiency for the horizontal muons, and large uncertainties in theoretical models in this area.}
	\label{fig:7}
\end{figure*}

For every bin in the zenith angle (see Figure~\ref{fig:6}) the starting energy for the spectra integration was chosen in such a way, that the integral of the theoretical muon flux becomes equal to the experimentally measured value. This energy is $E_{thr}$ since no muons below this starting energy contribute to the detected flux. The values of $E_{thr}$ determined by such an algorithm for the two variants of the surface spectra are shown in Figure~\ref{fig:7}. The area with $\cos\theta < 0.06$ is excluded from the further consideration due to the low detection efficiency for the horizontal muons, and large uncertainties in theoretical models in this area. Yet no notable influence on the result is expected because of the small number of horizontal muons.

Theoretical value of $\alpha$ and the effective temperature depend on an averaged value of $E_{thr}\cos\theta$. At the same time, theoretical value of $\beta$ independently relies on $E_{thr}$ and $\cos\theta$. These three parameters are separately calculated by the following formulae:
\begin{eqnarray}
	\langle E_{thr}\cos\theta\rangle &=& \sum_{i}E_{thr}\left(\cos\theta_i\right)n_i\cos\theta_i, \\
	\langle E_{thr} \rangle &=& \sum_{i}E_{thr}\left(\cos\theta_i\right)n_i, \\
	\langle \cos\theta \rangle &=& \sum_{i}n_i\cos\theta_i, 
\end{eqnarray}
\noindent
where index $i$ runs over all bins of each histogram in Figure~\ref{fig:6} with $\cos\theta$ $\geq$ 0.06, $\cos\theta_i$ – is the cosine value at the corresponding bin center and n$_i$ is the normalized muon flux – value of the corresponding bin given the total sum of all bins in a histogram is normalized to unity.

The resulting values of $\langle E_{thr}\cos\theta \rangle$, $\langle E_{thr}\rangle$ and $\langle \cos\theta \rangle$ for both spectra and the three detector positions are shown in Table~\ref{tab:1}. The spectrum \cite{gaisser} is among the most recently developed and the most commonly used one, so the best estimates in the right column are based on this spectrum. The numbers based on the other spectrum \cite{spec1} are used for the evaluation of the systematic errors caused by the theoretical uncertainty which are also indicated in the rightmost column of Table~\ref{tab:1}. Besides that, the systematic uncertainties due to the imperfect reconstruction of the angles were estimated. For the parameters $\langle E_{thr}\cos\theta \rangle$ and $\langle E_{thr}\rangle$ they turned out to be totally negligible compared to the uncertainties from the spectrum models. Yet for $\langle \cos\theta \rangle$ the influence of the angular distortion is much bigger than the statistical errors, so exactly these values are presented in Table~\ref{tab:1}.

The validity interval of equation~\ref{eq:1} defines the upper integration limit of $10^5$~GeV for both muon spectra when the threshold energy is calculated. Additional estimate of the systematic uncertainties was performed by changing this value to $10^4$~GeV. The procedure showed no changes in the calculated values down to the fifth decimal place.

\begin{table*}
\begin{center}
\caption{Averaged values of $\langle E_{thr}\cos\theta \rangle$, $\langle E_{thr}\rangle$ and $\langle \cos\theta \rangle$ for the two surface spectra and three detector positions}
\label{tab:1} 
\begin{tabular}{|c|c|c|c|}
	\hline
	\rule[-0.6em]{0cm}{1.6em}
	 & Spectrum \cite{spec1} & Spectrum \cite{gaisser} & Best estimate \\
	\hline
	\rule[-0.6em]{0cm}{1.6em}
	$\langle E_{thr}\cos\theta \rangle_{up}$ & 7.182$\pm$0.004 GeV & 8.032$\pm$0.004 GeV & 8.0$\pm$0.9 GeV \\
	\hline
	\rule[-0.6em]{0cm}{1.6em}
	$\langle E_{thr}\cos\theta \rangle_{mid}$ & 6.983$\pm$0.004 GeV & 7.815$\pm$0.004 GeV & 7.8$\pm$0.8 GeV \\
	\hline
	\rule[-0.6em]{0cm}{1.6em}
	$\langle E_{thr}\cos\theta \rangle_{down}$ & 6.833$\pm$0.004 GeV & 7.664$\pm$0.004 GeV & 7.7$\pm$0.8 GeV \\
	\hline
	\rule[-0.6em]{0cm}{1.6em}
	$\langle E_{thr}\rangle_{up}$ & 11.12$\pm$0.03 GeV & 12.81$\pm$0.03 GeV & 12.8$\pm$1.7 GeV \\
	\hline
	\rule[-0.6em]{0cm}{1.6em}
	$\langle E_{thr}\rangle_{mid}$ & 10.79$\pm$0.04 GeV & 12.45$\pm$0.04 GeV & 12.5$\pm$1.7 GeV \\
	\hline
	\rule[-0.6em]{0cm}{1.6em}
	$\langle E_{thr}\rangle_{down}$ & 10.54$\pm$0.03 GeV & 12.19$\pm$0.03 GeV & 12.2$\pm$1.7 GeV \\
	\hline
	\rule[-0.6em]{0cm}{1.6em}
	$\langle \cos\theta \rangle_{up}$ & \multicolumn{3}{c|}{0.656$\pm$0.007} \\
	\hline
	\rule[-0.6em]{0cm}{1.6em}
	$\langle \cos\theta \rangle_{mid}$ & \multicolumn{3}{c|}{0.654$\pm$0.008} \\
	\hline
	\rule[-0.6em]{0cm}{1.6em}
	$\langle \cos\theta \rangle_{down}$ & \multicolumn{3}{c|}{0.655$\pm$0.008} \\
	\hline
\end{tabular}
\end{center}
\end{table*}

\section{Correlation effects}

\subsection{Atmospheric data}
The data on the atmospheric temperature and pressure were obtained from ERA5 \cite{era5}, one of the databases of European Center for Medium-Range Weather Forecasts (ECMWF). The database uses several kinds of data sources, like measurements from the ground level stations, balloons and satellites, coming from the entire world. The final global data are obtained from the extrapolation of these measurements. This work considers the temperature values at 37 different pressure levels, from 1 mbar to 1000 mbar, and the values of the pressure on the ground level. The data was taken for the DANSS location (57.9$^o$ N, 35.1$^o$ E) on an hourly basis.

For the direct analysis of the temperature influence on the muon flux, the temperature at the point of muon production must be known. This is impossible, but instead an integral approach can be used, for example the isothermal approximation of the atmosphere~\cite{Kpi} with the effective temperature equal to:

\begin{equation}
	\label{t_eff}
	T_{eff}=\frac{\int^\infty_0dX~T(X)W(X)}{\int_{0}^{\infty}dX~W(X)}\simeq\frac{\sum_i\Delta X_iT(X_i)W(X_i)}{\sum_i\Delta X_iW(X_i)}
\end{equation}

\noindent
where $X$ – is the atmospheric depth in g/cm$^2$, which could be got from the equation: 1 mbar = 1.019 g/cm$^2$, $W(X)$ – is a weight function, which shows how much the pressure level $X_i$  impacts on the muon production. In the recent approach~\cite{Kpi}, $W(X)$ consists of two terms: $W_\pi(X)$ and $W_K(X)$ – the functions of the pion and kaon contributions, accordingly. Although for our muon threshold energy range the contribution of kaons is very low, it is taken into account in this research, and each of the functions can be written as:

\begin{equation}
	\label{eq:8} 
	W_{\pi,K}(X)\simeq\frac{\left(1-\frac{X}{\lambda_{\pi,K}}\right)^2e^{-\frac{X}{\Lambda_{\pi,K}}}A_{\pi,K}}{\gamma+(\gamma+1)B_{\pi,K}K(X)\left(\frac{\langle E_{thr}\cos\theta\rangle}{\epsilon_{\pi,K}}\right)^2},
\end{equation}

\begin{equation}
	\label{eq:9} 
	K(X)\equiv\frac{X\left(1-\frac{X}{\lambda_{\pi,K}}\right)^2}{\left(1-e^\frac{X}{\lambda_{\pi,K}}\right)\lambda_{\pi,K}},
\end{equation}

\begin{equation}
	\label{eq:10} 
	\frac{1}{\lambda_{\pi,K}}=\frac{1}{\Lambda_N}-\frac{1}{\Lambda_{\pi,K},}
\end{equation}

\noindent
$\Lambda_{N,\pi,K}$ is the interaction length in the atmosphere of the primary cosmic nucleons and secondary pions and kaons, accordingly; $A_{\pi,K}$ is a constant which includes masses of muons and mesons, muon spectral index and kinematic parameters; $B_{\pi,K}$ takes into consideration the relative attenuation of the mesons; $\gamma$ is the muon spectral index, and $\epsilon$$_{\pi,K}$ is the critical energy of the mesons at which they decay or interact with equal probability. All corresponding numeric values are given in Table~\ref{tab:2}.

\begin{table}
	\caption{Parameter values and their errors used in equations~\ref{eq:8}-\ref{eq:10} and~\ref{eq:3}.}
	\label{tab:2}
	\begin{tabular}{ccc}
		\hline\noalign{\smallskip}
		Parameter & Value & References\\
		\noalign{\smallskip}\hline\noalign{\smallskip}
		A$_\pi$ & 1 & \cite{minosfar} \\
		A$_K$ & 0.38r$_K/\pi$ & \cite{minosfar} \\
		r$_{K/\pi}$ & 0.149$\pm$0.06 & \cite{gaisser} \\
		B$_\pi$ & 1.460$\pm$0.007 & \cite{minosfar} \\
		B$_K$ & 1.740$\pm$0.028 & \cite{minosfar} \\
		$\Lambda$$_N$ & 120 g/cm$^2$ &  \cite{gaisser} \\
		$\Lambda$$_\pi$ & 180 g/cm$^2$ & \cite{gaisser} \\
		$\Lambda$$_K$ & 160 g/cm$^2$ & \cite{gaisser} \\
		$\gamma$ & 1.7$\pm$0.1 & \cite{minosmuon} \\
		$\epsilon$$_\pi$ & 114$\pm$3 GeV & \cite{minosfar} \\
		$\epsilon$$_K$ & 851$\pm$14 GeV & \cite{minosfar}  \\
		\noalign{\smallskip}\hline
	\end{tabular}
\end{table}

\begin{figure}
	\includegraphics[width=0.49\textwidth]{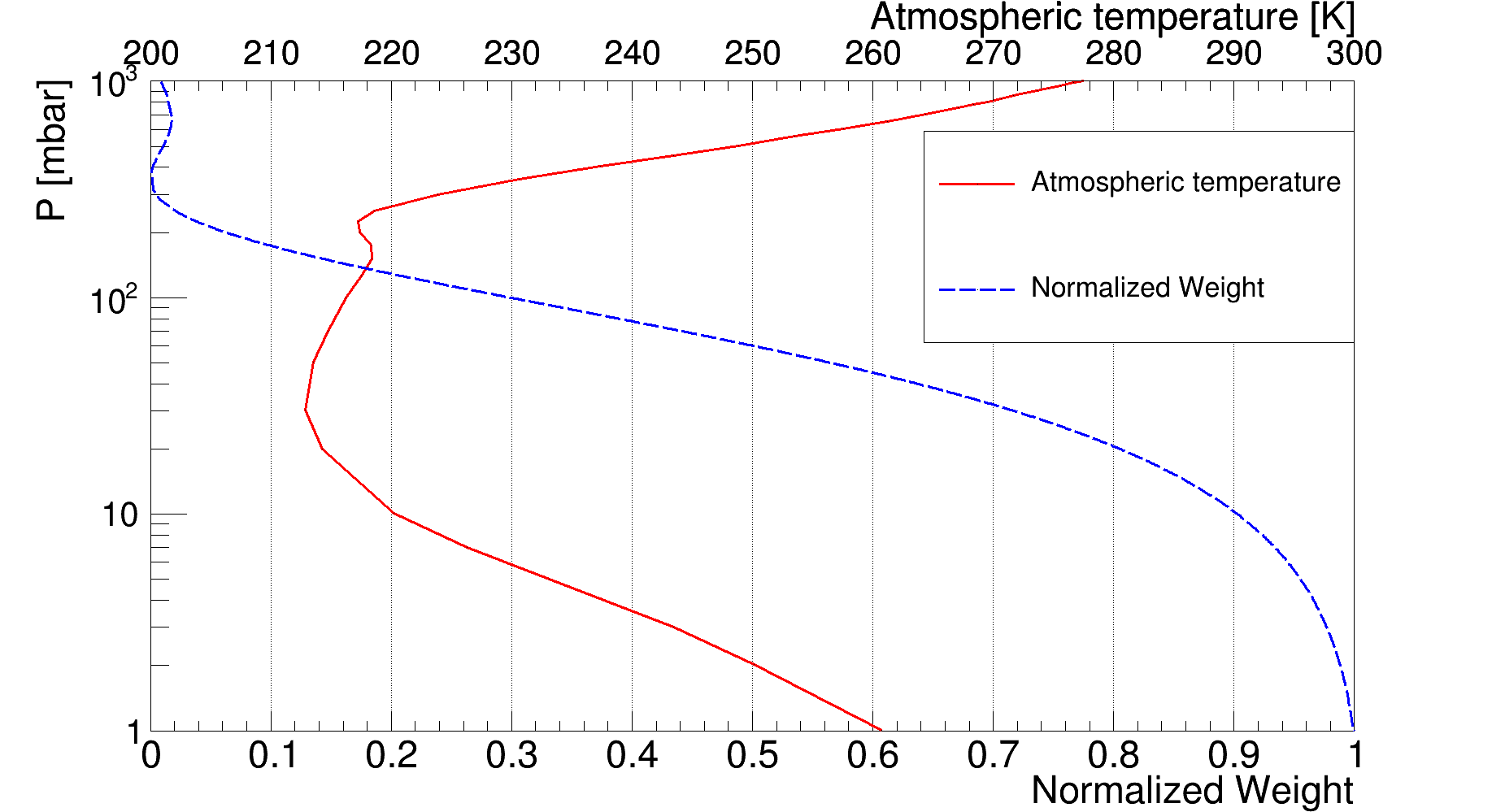}
	\caption{Normalized weight function W (dashed blue line) and the average temperature T (solid red line) at different heights (pressure levels).}
	\label{fig:8}
\end{figure}

\begin{figure}
	\includegraphics[width=0.49\textwidth]{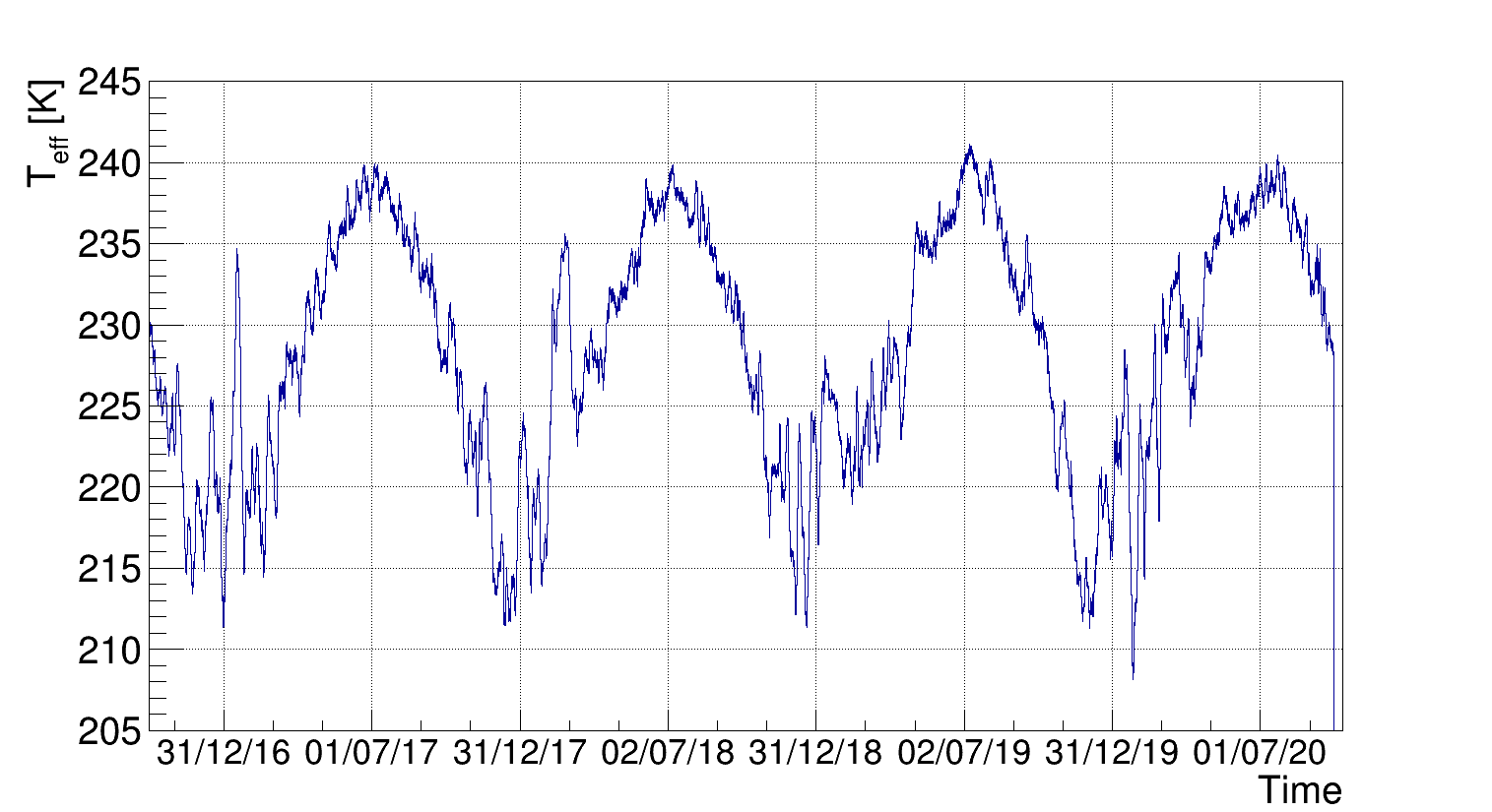}
	\caption{Effective atmospheric temperature versus the calendar time}
	\label{fig:9}
\end{figure}

\begin{figure}
	\includegraphics[width=0.49\textwidth]{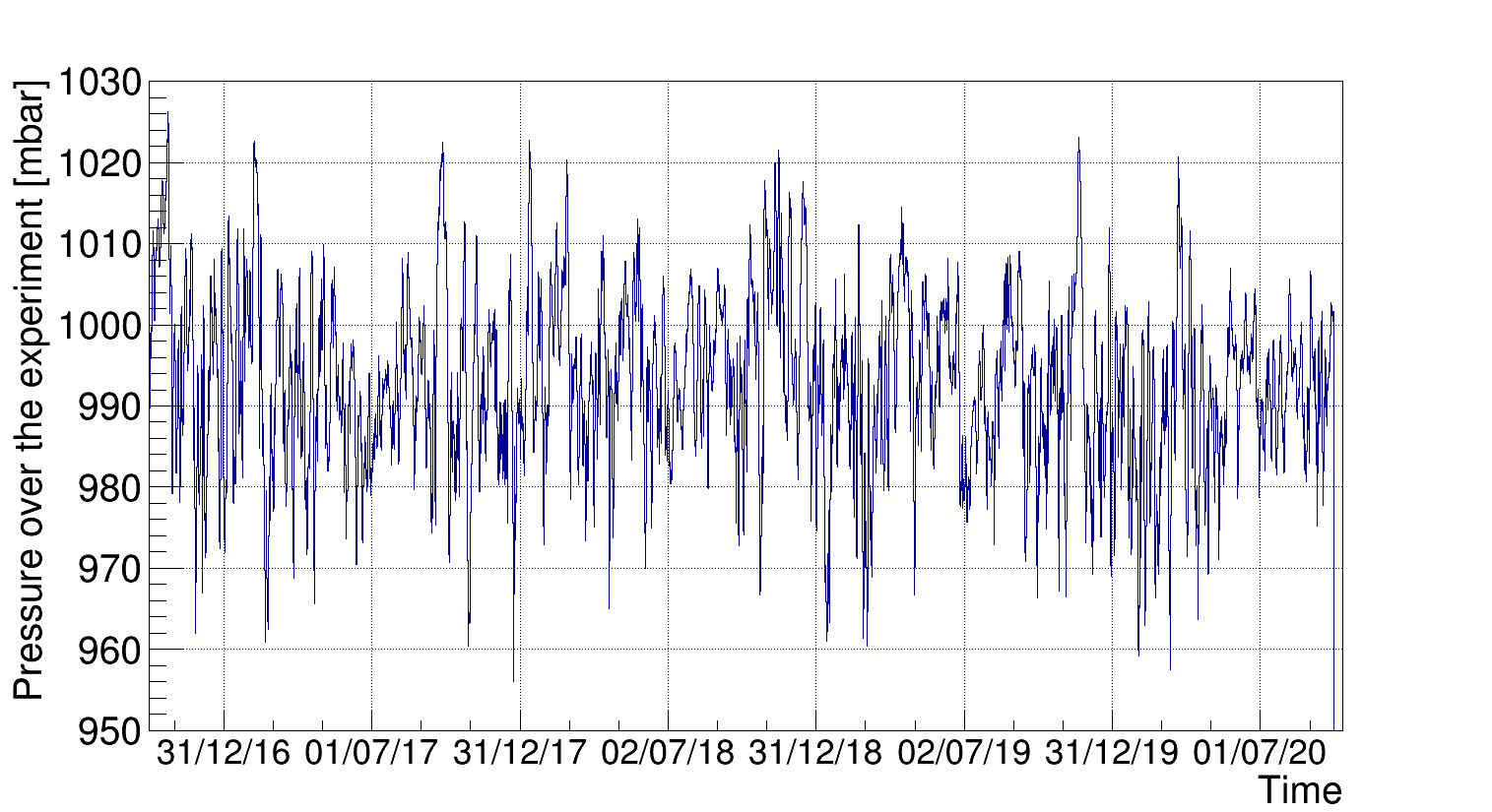}
	\caption{Atmospheric pressure on the ground level versus the calendar time.}
	\label{fig:10}       
\end{figure}

The behavior of the normalized $W(X)$ function and of the average temperature $T(X)$ upon the height above the ground, or, equivalently, on the pressure level are shown in Figure~\ref{fig:8}. The dependencies of the effective temperature and of the ground level pressure over the time estimated for the location of the DANSS detector are shown in Figures~\ref{fig:9} and~\ref{fig:10}. By comparison of Figure~\ref{fig:9} and Figure~\ref{fig:4} it becomes clear that the expected temperature effect should be negative.

There are no meteorological stations making regular measurements near the power plant, so there is no direct possibility to estimate the error of the temperature calculations at the DANSS location. Yet to evaluate this error the experimentally measured temperature from Bologoye meteorological station \cite{igra} (about 60 km West of KNPP) was compared to the predictions of ERA5 model for this location. The distribution width of the difference between the model predictions and the station data is $\sigma T_{eff}$=0.94 K. In addition the average difference is shifted from zero by $\Delta T_{eff}$=0.12~K. To estimate the pressure uncertainties the data from ERA5 was compared to that from the local weather archive \cite{localarchive} during one calendar year of 2018. The resulting pressure error was $\sigma P$=59~Pa. $\sigma T_{eff}$ and $\sigma P$ are accounted for as individual errors for each hourly value of the effective temperature and pressure (see Figure~\ref{fig:11}). The influence of the shifts will be discussed later in this paper.

\subsection{Correlation analysis}
	The goal of this research is to determine how the cosmic muon flux relates to the atmospheric temperature and pressure. In this work, the parameters characterizing these dependencies are the correlation coefficients between the relative variation of the muon rate and the relative change of the effective temperature ($\alpha$) or the absolute deviation of the ground level pressure ($\beta$), both taken in comparison to their average values:
	\begin{equation}
		\label{separate_effects}
		\frac{I-\langle I\rangle}{\langle I\rangle}=\alpha\frac{T_{eff}-\langle T_{eff}\rangle}{\langle T_{eff}\rangle};~\frac{I-\langle I\rangle}{\langle I\rangle}=\beta\left(P-\langle P\rangle\right)
	\end{equation}
	These definitions are chosen to simplify the comparison of the results on $\alpha$ and $\beta$ to the theoretical predictions by Barrett \cite{barret} and Sagisaka \cite{sagisaka1986}. For the direct comparison we use the mean values of the effective temperature, of the ground level pressure and of the muon rate calculated for each detector position separately (see Table~\ref{tab:3}).
	
	\begin{table}
		\caption{Mean values of the muon rate, of the effective temperature and of the ground level pressure for each detector position.}
		\label{tab:3}       
		\begin{tabular}{llll}
			\hline\noalign{\smallskip}
			 & $\langle$I$\rangle$ [muons/s] & $\langle$T$_{eff}$$\rangle$ [K] & $\langle$P$\rangle$ [mbar] \\
			\noalign{\smallskip}\hline\noalign{\smallskip}
			Up position & 13.219$\pm$0.001 & 228.85$\pm$0.08 & 992.8$\pm$0.1 \\
			Middle position & 13.664$\pm$0.002  & 226.70$\pm$0.10 & 994.0$\pm$0.2 \\
			Down position & 14.024$\pm$0.001 & 227.45$\pm$0.07 & 992.8$\pm$0.1 \\
			\noalign{\smallskip}\hline
		\end{tabular}
	\end{table}
	
	The differences in the mean values of $\langle T_{eff}$$\rangle$ and $\langle P\rangle$ are due to the heterogeneity of the exposure periods in each of the detector positions. For example, the middle position was excluded from data taking after April 2019, and thus the corresponding data set misses the warm part of the third year. In addition due to various reasons the detector spent about two months in the down position in the winter 2019$^{th}$ and several months in the up position in the spring and at the start of the summer 2020$^{th}$.

\begin{figure}
		\begin{minipage}[h]{1\linewidth}
			\center{\includegraphics[width=1\linewidth]{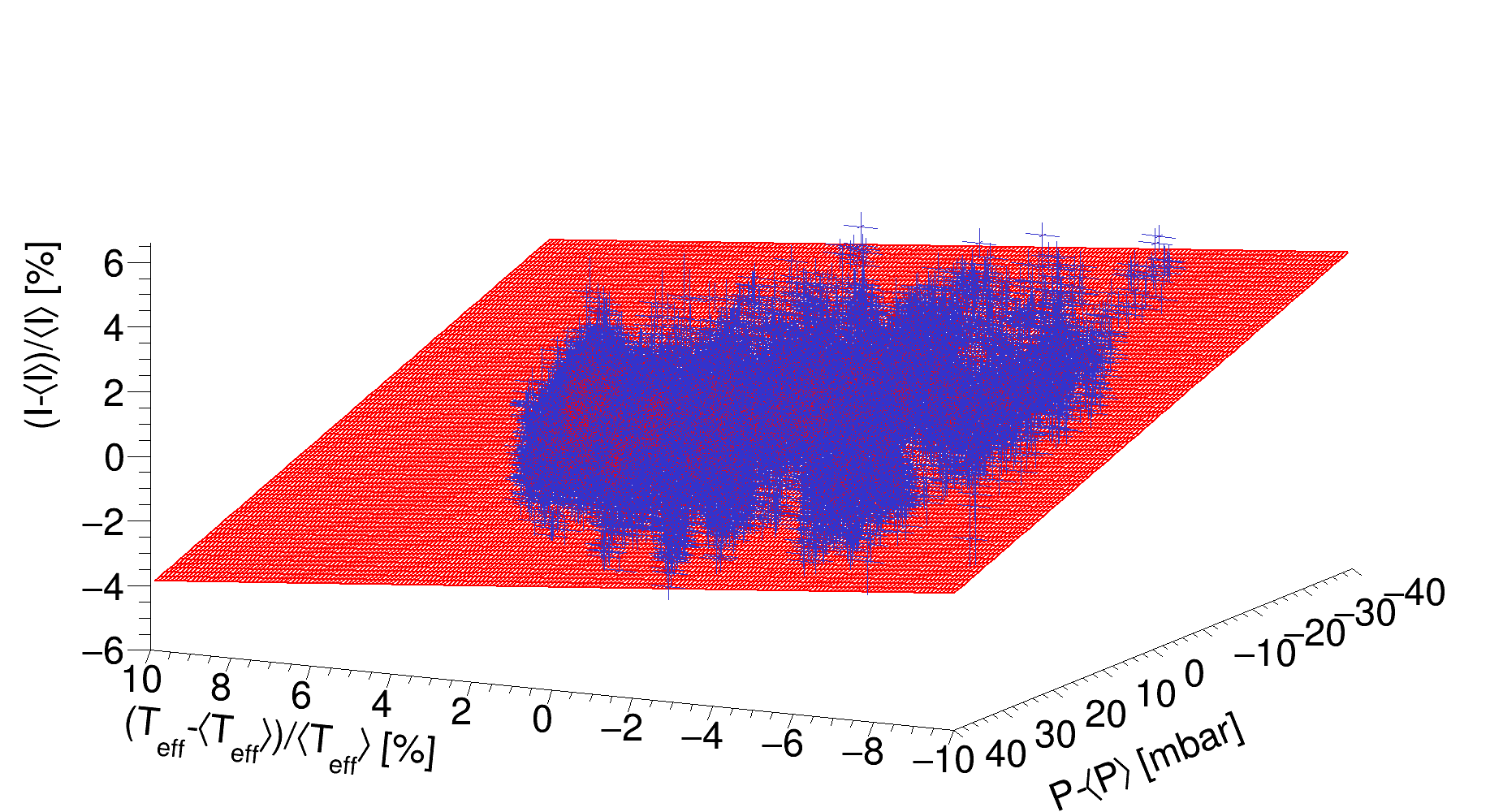} \\ a)}
		\end{minipage}
		\vfill
		\begin{minipage}[h]{1\linewidth}
			\center{\includegraphics[width=1\linewidth]{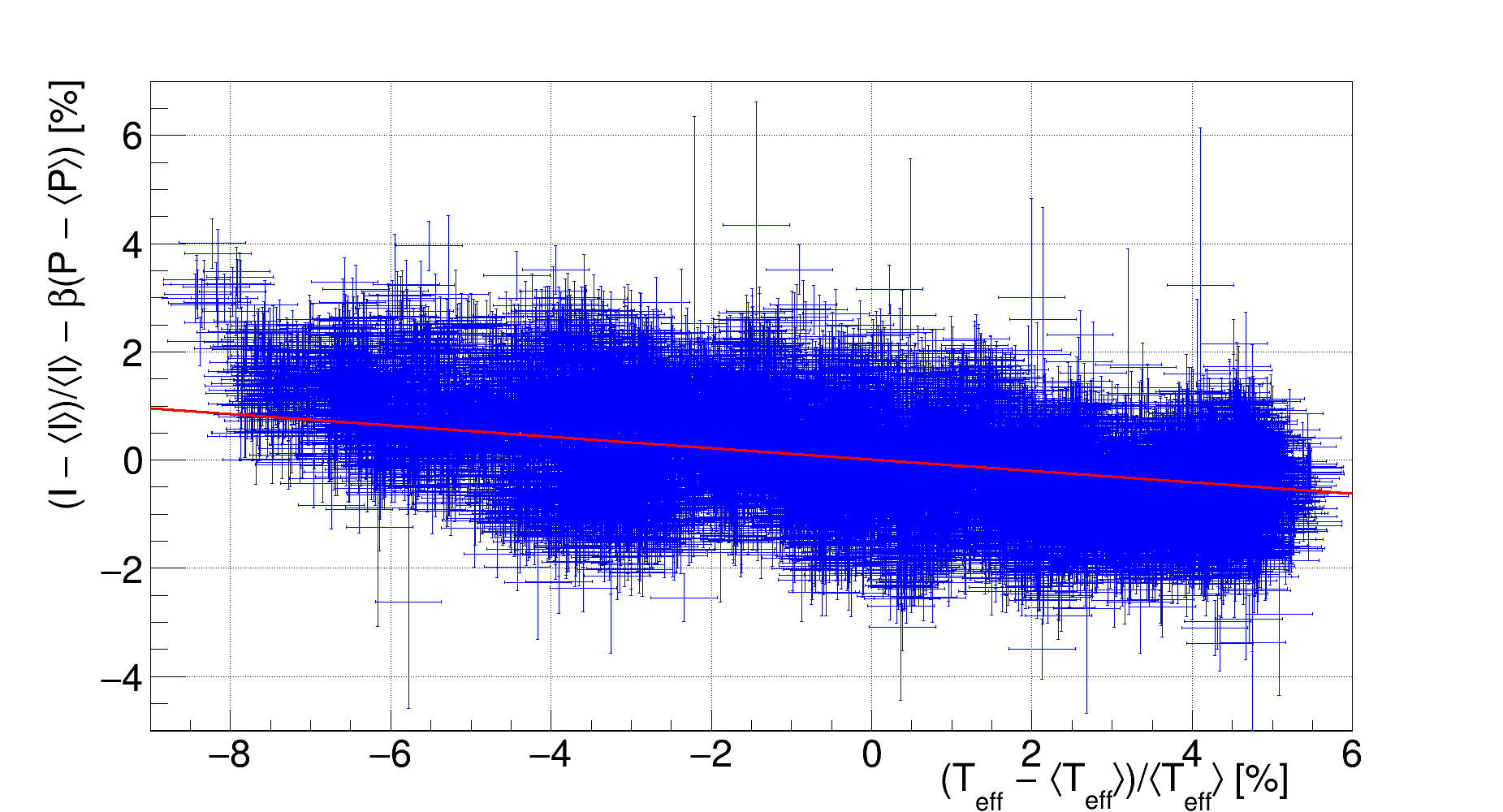} \\ b)}
		\end{minipage}
		\vfill
		\begin{minipage}[h]{1\linewidth}
			\center{\includegraphics[width=1\linewidth]{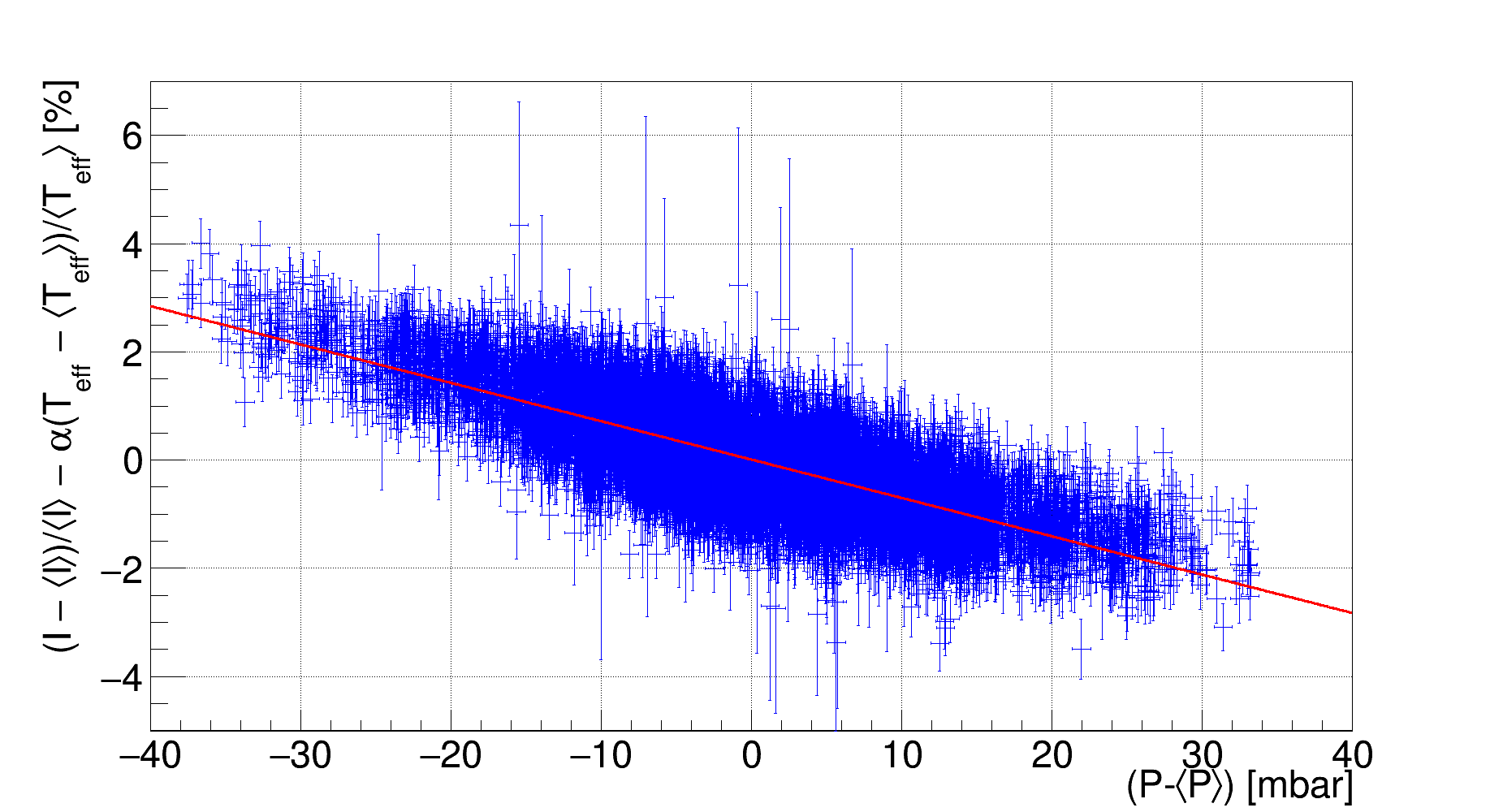} \\ c)}
		\end{minipage}	
		\caption{Relative variation of the muon flux vs the relative variation of the effective	atmospheric temperature and vs the absolute variation of the surface pressure for the down detector position.}
		\label{fig:11}
	\end{figure}

	The value of the coefficient $\alpha$ is expected to be small. This is due to the intermediate position of DANSS in terms of the muon threshold energy, where positive and negative temperature effects are of similar magnitude. Accordingly, the temperature correlations would be significantly blurred by the barometric effect, which is not expected to be suppressed, and, in turn, this effect will lead to big uncertainties in $\alpha$ calculation. To avoid this difficulty and simultaneously accurately take into account both effects a three-dimensional description of the muon flux as a function of both the temperature and the pressure was introduced:
	\begin{equation}
		\label{sum_effect}
		\frac{I-\langle I\rangle}{\langle I\rangle}=\alpha\frac{T_{eff}-\langle 	T_{eff}\rangle}{\langle T_{eff}\rangle}+\beta\left(P-\langle P\rangle\right)+c
	\end{equation}
	It is clear that $\alpha$ and $\beta$ in equation~\ref{sum_effect} have the same meanings as in equations~\ref{separate_effects}. Indeed, $\frac{I-\langle I\rangle}{\langle I\rangle}=\alpha\frac{T_{eff}-\langle T_{eff}\rangle}{\langle T_{eff}\rangle}$ is the equation of the line obtained when the plane in eq.~\ref{sum_effect} crosses the plane $P-\langle P\rangle$=0, and similar reasoning is true for the barometric coefficient. The constant term $c$ is expected to be zero. The whole data set was fit using this three-dimensional approach. An example of such fit for the detector in down position is shown in Figure~\ref{fig:11}. The central and bottom panels of the figure illustrate the muon flux correlations with the individual atmospheric parameters in a 'purified' way. In the panel b) the data in the flux-temperature plane is corrected for the influence of the pressure deviations using the value of $\beta$-coefficient, obtained from the fit. Similarly, the panel c) shows the pressure correlation for the data with the compensated temperature effect. The numerical results of the fits for the three detector positions are summarized in the columns 2 and 3 of Table~\ref{tab:4}. The fit constant $c$ is consistent with zero for all three cases. The example in Figure~\ref{fig:11} gives an idea of the remaining value of this term for the down detector position.
	
\subsection{Systematic uncertainties}
	Three sources of the systematic uncertainties are considered in this research.
	The first source arises from the ERA5 prediction uncertainty (see section 4.1). $\sigma T_{eff}$ and $\sigma P$ automatically appear in the resulting statistical errors because both are accounted by the fit in Figure~\ref{fig:11}. The temperature shift $\Delta T_{eff}$ is negligible compared to the corresponding average value $\langle T_{eff}\rangle$ and the same is true about its contribution to the systematic error. The pressure shift of any value can not at all affect the correlation coefficients since the averaged pressure value gets shifted by the same magnitude as every individual pressure value.

	Following \cite{minosnear} the next source is attributed to the inaccuracy of the integration procedure in $T_{eff}$ calculations. The values in the numerator of Eq. (\ref{t_eff}) depend on the integration method. To evaluate the corresponding errors the correlation coefficients were estimated using several numerical approaches to the integration procedure. While the trapezoidal rule was chosen as the basic integration method, the maximum differences between several tested methods were taken as estimates of the uncertainty from integration. Corresponding numbers could be observed in the 4$^{th}$ column of Table~\ref{tab:4}.

	The third contribution to the uncertainty comes from the errors of the parameters in Table~\ref{tab:1} and of the measured values in Table~\ref{tab:2}, both used in $T_{eff}$ computation. Most of the above have effectively no impact on $\alpha$ and $\beta$, and only errors in $\langle E_{thr}\cos\theta \rangle$ and $r_{K/\pi}$ have small influence on the correlation coefficients. Corresponding numbers can be found in the 5$^{th}$ column of Table~\ref{tab:4}.

	The total systematic uncertainty is presented in the last column of Table~\ref{tab:4}. The values of the correlation coefficients in the three detector positions are very close to each other, yet they differ beyond the range of the estimated uncertainties. Besides, the detection conditions, such as the muon rate and the threshold energy distributions, are quite different in various positions. This is the reason for keeping three separate values for the correlation coefficients rather than average them over the detector positions.

	\begin{table*}
		\begin{center}
			\caption{The resulting values and the error estimates for the temperature and barometric correlation coefficients for the each detector position.}
			\label{tab:4} 
			\begin{tabular}{cccccc}
				\hline\noalign{\smallskip}
				\multirow{2}{*}{Parameter} & \multirow{2}{*}{Value} & \multirow{2}{*}{Stat. uncertainty} & \multicolumn{3}{c}{Sys. uncertainty} \\
				& & & Integral & Parameters & Total \\
				\noalign{\smallskip}\hline\noalign{\smallskip}
				$\alpha_{up}$ & $-0.0861$ & 0.0014 & 0.00026 & 0.00023 & 0.0003 \\
				$\alpha_{mid}$ & $-0.0904$ & 0.0021 & 0.00140 & 0.00018 & 0.0014 \\
				$\alpha_{down}$ & $-0.1051$ & 0.0012 & 0.00098 & 0.00022 & 0.0010 \\
				$\beta_{up} [\%/mbar]$ & $-0.0669$ & 0.0004 & 0.000071 & 0.000013 & 0.0001 \\
				$\beta_{mid} [\%/mbar]$ & $-0.0677$ & 0.0006 & 0.000380 & 0.000013 & 0.0004 \\
				$\beta_{down} [\%/mbar]$ & $-0.0709$ & 0.0004 & 0.000068 & 0.000028 & 0.0001 \\
				\noalign{\smallskip}\hline\noalign{\smallskip}
			\end{tabular}
		\end{center}
	\end{table*}

\subsection{Comparison with theory predictions and other experiments}
	One can see from Figure~\ref{fig:7} that the threshold energy in the vertical and horizontal directions significantly deviates from its mean value, determined mostly by the flat region in the center. And it seems interesting to separately evaluate the correlation effects in these areas of zenith angle. So the same analysis was repeated for the two angular sub-ranges: nearly horizontal region with $\cos\theta < 0.36$, and nearly vertical region with $\cos\theta > 0.9$. The results for these angular ranges are given in Table~\ref{tab:5} together with those for the whole upper hemisphere.
	
	The theoretical predictions \cite{barret} for $\alpha$ in case of the low $E_{thr}$ values consist of two nonnegligible terms, called positive ($\delta$) and negative ($\delta'$) temperature effects:
	\begin{equation}
		\alpha=\delta-\delta',
	\end{equation}
	\begin{equation}
		\delta=1/\left[\frac{\gamma}{\gamma+1}\frac{\epsilon_\pi}{1.1\langle E_{thr}\cos\theta\rangle}+1\right],
	\end{equation}
	\begin{equation}
		\delta'=\left(\frac{1}{E_{thr}\cos\theta}\right)\left(\frac{m_\mu c^2H\gamma}{c\tau_\mu(\gamma+1)}\right)ln\left(\frac{1030}{\Lambda_N\cos\theta}\right),
	\end{equation}
	where $\tau_\mu$ is the muon mean life time, $H=RT/Mg$ with $M$ – the molecular weight of air, $g$ – gravitational acceleration, $R$ – gas constant, $T$ – atmospheric temperature. Average value of $T$ and, consequently, of $H$ in the area of muon production is practically independent of the height, so $H$ is considered constant with the value $H=6.46\cdot10^5$~cm. Since $\delta'$ depends not only on the average parameter $\langle E_{thr}\cos\theta\rangle$, but separately on the muon threshold energy and the zenith angle cosine, it was calculated in the following way: for each value of $\langle E_{thr}\cos\theta\rangle$ in the range 3--1900 GeV the distributions from Figure~\ref{fig:7}b were scaled to represent this average value. These scaled distributions were then used to calculate the average magnitude of $\delta'$ at corresponding $\langle E_{thr}\cos\theta\rangle$.

\begin{figure*}
	\centering
		\includegraphics[width=0.99\textwidth]{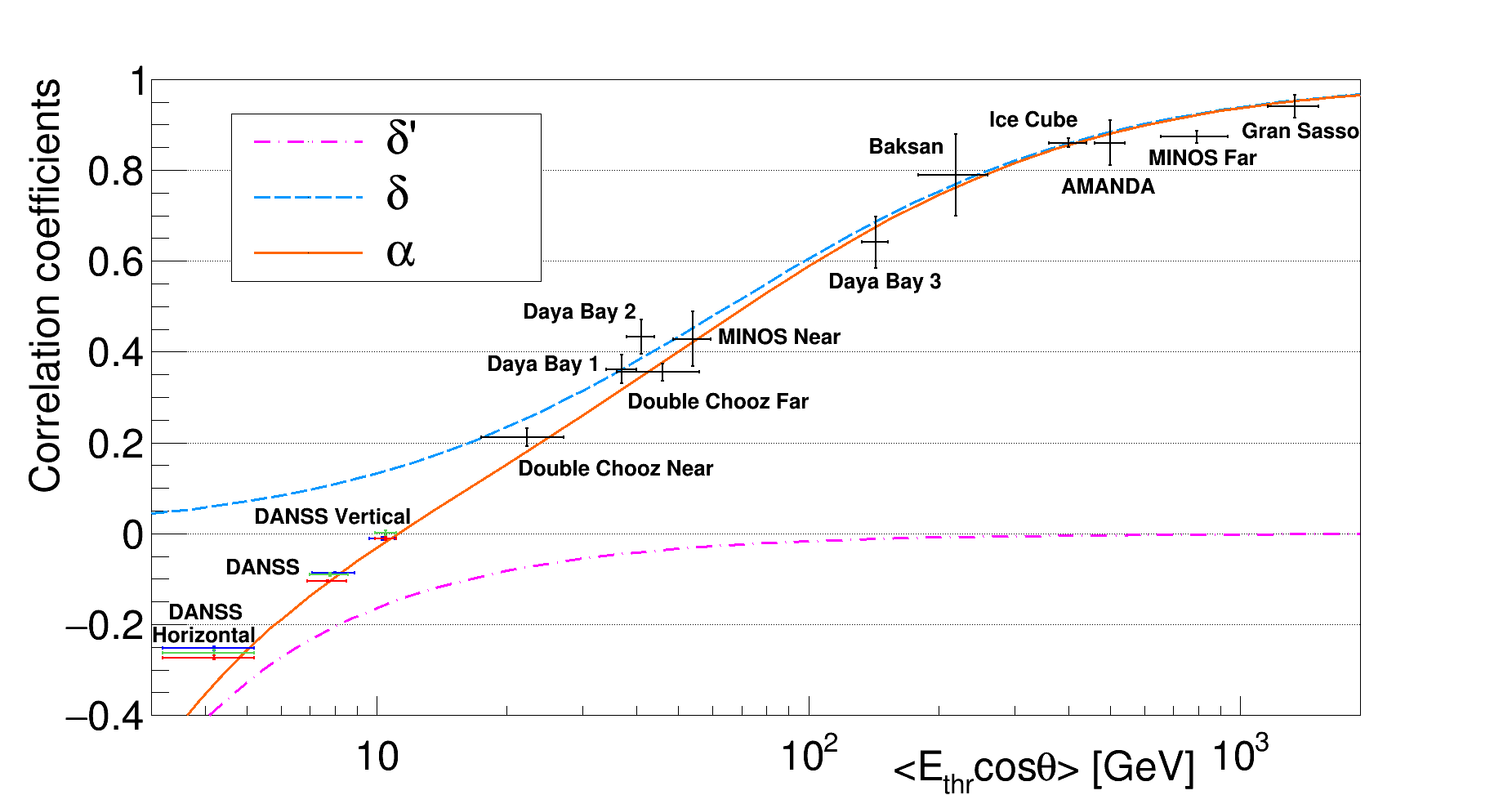}
		\caption{Comparison of the temperature correlation coefficient from different experiments to the model curves. The blue dashed line is for the theoretical positive term $\delta$, the negative term $(-\delta')$ is shown in dash-dotted pink, while the orange solid line designates the total temperature correlation coefficient $\alpha$. Three groups of the colored markers represent the DANSS results for the two angular sub-ranges ('Vertical' and 'Horizontal') and for the whole sky (no additions). The three colors denote the detector position: red means down, green and blue are for middle and up positions, correspondingly. The results from the other experiments come from Double Chooze Near and Far \cite{doublechooz} cites, from MINOS Near \cite{minosnear} and Far \cite{minosfar} detectors, as well as from the three experimental halls of Daya Bay \cite{dayabay} and avereged result of the experiments located in Gran Sasso, including: LVD \cite{LVD}, MACRO \cite{macro}, Borexino \cite{borexino}, GERDA \cite{gerda} and OPERA \cite{opera}. To compare results with AMANDA \cite{amanda}, IceCube \cite{icecube} and Baksan \cite{baksan} their values of $\langle E_{thr}\cos\theta\rangle$ are taken from work \cite{dayabay}.}
		\label{fig:12}  
	\end{figure*}

	The curves that present the theoretical predictions for the positive, negative and summary effects together with the experimental results of this work and that of some other experiments are shown in Figure~\ref{fig:12}. The values from DANSS are close to zero or have negative sign, in contrast to the other measurements. Yet the DANSS results are in perfect agreement with the theory predictions for all three positions of the detector and all three angular areas. The results of some experiments like Hobart \cite{hobart}, Matsushiro \cite{matsushiro} and Poatina \cite{poatina} are not present in the comparison. They use approaches other than the effective temperature method, and their results can not be directly compared to the others in Figure~\ref{fig:12}.
	
	The theoretical dependence of the correlation coefficient $\beta$ on $E_{thr}$ and $\cos\theta$ is more complicated. In the area of the threshold energy values typical for the DANSS location the total barometric coefficient is a sum of three essential terms:
	\begin{equation}
		\beta=\beta_a+\beta_d+\beta_p,
	\end{equation}
	where $\beta_a$ reflects the absorption by the ionization loss, $\beta_d$ describes the growth of the muon decays and $\beta_p$ represents the increase of the muon production. All these three terms depend separately on the threshold energy, on the zenith angle and on the measurement altitude, and in much more complicated way than the temperature effects. So while comparing to the experimental results it is common to draw the theoretical dependencies of $\beta(E_{thr}$) with fixed values of the altitude and of the zenith angle. For more details about the barometric effect see \cite{sagisaka1986}.
	
\begin{figure}
		\includegraphics[width=0.49\textwidth]{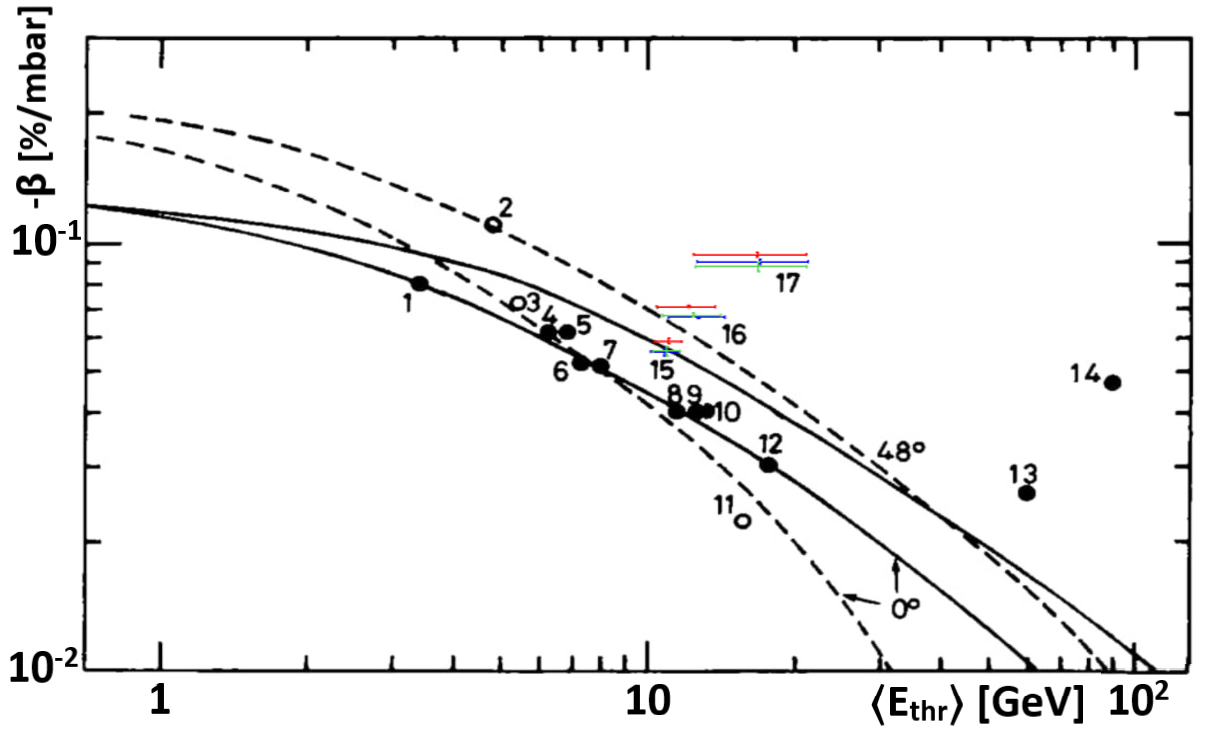}				\caption{Comparison of the experimental values of the barometric correlation coefficients $\beta$ measured at different stations with model calculations for X$_0$=1000 g/cm$^2$ (solid line) and X$_0$=600 g/cm$^2$ (dashed line); the lower curves are for 0$^o$ and the upper curves are for 48$^o$ of the zenith angle. The open and solid circles represent the stations located higher and lower than 1000 meters, accordingly. The results of this work are designated by numbers: 15 for nearly vertical muons, 16 for the total muon flux, and 17 for almost horizontal muons. The down position of DANSS is marked by red; middle and up positions have green and blue colors. The other stations are: 1) Yakutsk (20 m.w.e. in depth), 2) Bolivia, 3) Embudo, 4) Mawson, 5) Misato, 6) Hobart, 7) Budapest, 8) Takeyama, 9) London , 10) Yakutsk (60 m.w.e. in depths), 11) Socoro, 12) Sakashita, 13) Matushiro \cite{matsushiro} and 14) Poatina \cite{poatina}. References for stations 1-12 can be found in \cite{else}.}
		\label{fig:13}  
	\end{figure}

	In Figure~\ref{fig:13} the results from DANSS are shown superimposed on the plots replicated from \cite{sagisaka1986}. The figure presents the barometric correlation coefficients observed at different stations, and the theory curves for certain values of the zenith angle and of the measurement altitude. The DANSS detector is located at approximately 175 meters above the sea level, which is very close to $X_0$=1000 g/cm$^2$ presented by solid lines. The mean values of the zenith angle $\theta$ for the analysis of the total muon flux lay near 48$^o$, which is typical for the experiments integrating their data over the whole sky. Thus the DANSS results in this case are directly comparable to the upper solid line in the figure. The results for nearly vertical muons should be compared to the lower solid line. The data for the muons close to the horizon are also plotted in the figure, but there is no model curve for this case. One can notice that all values of the barometric correlation coefficients measured in this work are approximately 30\% above the model predictions and the discrepancy is far beyond the estimated uncertainty.

\section{Conclusion}
	Measurements of the temperature and barometric correlation coefficients have been performed by the DANSS detector based on the four years dataset. The distinctive feature of this work is the unique position of the muon detector: though located even above the ground surface, it is significantly screened by a massive reactor core surrounded by water pools and tons of concrete. The advantage of the lifting system to change the distance between the detector and the reactor core gives an opportunity to reveal the fine difference in the correlation coefficients for slightly different muon angular distributions and total fluxes.
	
	The observed numerical values are summarized in Table~\ref{tab:5}. Along with the results based on the total muon flux, the magnitude of the effects was estimated for two angular sub-ranges, where the overburden parameters are considerably different -- the areas of horizontal and vertical angles. The results from the three detector positions differ significantly if compared to the combination of the statistical and systematic errors. The temperature coefficient $\alpha$ is in good agreement with theoretical predictions for all the three detector positions and in all angular ranges. Yet the sign of the effect is negative and opposite to the most of the other experiments because of the specific location of the DANSS detector. The barometric coefficient $\beta$ is measured noticeably higher than the theoretical prediction in the angular intervals where the comparison is straightforward. The difference is significant compared to the estimated uncertainties. Together with the results from some other groups this may be an evidence of the imperfection of the theoretical approaches to the barometric effect. The team of the DANSS experiment hopes that the discrepancy of our result with the theoretical predictions will inspire searches for the explanation of this phenomena and give a new impulse to the development of the cosmic rays models.
	\begin{table*}
	\begin{center}
		\caption{Summary of the DANSS results on the temperature and barometric correlation coefficients.}
		\label{tab:5} 
		\begin{tabular}{c|c|c|c|c|c|c}
			\hline
			\multirow{2}{*}{Angular region} & Detector & Correlation & \multirow{2}{*}{Experimental value} & $\langle E_{thr}\cos\theta$$\rangle$ & $\langle E_{thr}$$\rangle$ & \multirow{2}{*}{$\langle\cos\theta$$\rangle$}\\
			& position & coefficient & & [GeV] & [GeV] & \\
			\hline

			\multirow{7}{*}{All angles}
			& \multirow{2}{*}{Up} & $\rule{0cm}{1.1em}\alpha$ & $-0.0861\pm$0.0014(stat.)$\pm$0.0003(sys.) & \multirow{2}{*}{8.0$\pm$0.9} & \multirow{2}{*}{12.8$\pm$1.7} & \multirow{2}{*}{0.656$\pm$0.007}\\
			& & $\rule[-0.6em]{0cm}{1.0em}\beta~[\%/mbar]$ & $-0.0669\pm$0.0004(stat.)$\pm$0.0001(sys.) &  &  &\\
			\cline{2-7}
			& \multirow{2}{*}{Middle} & $\rule{0cm}{1.2em}\alpha$ & $-0.0904\pm$0.0021(stat.)$\pm$0.0014(sys.) & \multirow{2}{*}{7.8$\pm$0.8}  & \multirow{2}{*}{12.5$\pm$1.7} & \multirow{2}{*}{0.654$\pm$0.008}\\
			& & $\rule[-0.6em]{0cm}{1.0em}\beta~[\%/mbar]$ & $-0.0677\pm$0.0006(stat.)$\pm$0.0004(sys.) &  & &\\
			\cline{2-7}	
			& \multirow{2}{*}{Down} & $\rule{0cm}{1.2em}\alpha$ & $-0.1051\pm$0.0012(stat.)$\pm$0.0010(sys.) & \multirow{2}{*}{7.7$\pm$0.8} & \multirow{2}{*}{12.2$\pm$1.7} & \multirow{2}{*}{0.655$\pm$0.008}\\
			& &\rule[-0.6em]{0cm}{1.0em} $\beta~[\%/mbar]$ & $-0.0709\pm$0.0004(stat.)$\pm$0.0001(sys.) &  & &\\
			\hline
			
			\multirow{7}{*}{Vertical}
			& \multirow{2}{*}{Up} & $\rule{0cm}{1.1em}\alpha$ & $-0.0111\pm$0.0032(stat.)$\pm$0.0009(sys.) & \multirow{2}{*}{10.3$\pm$0.6} & \multirow{2}{*}{10.9$\pm$0.7} & \multirow{2}{*}{0.950$\pm$0.002}\\
			& & $\rule[-0.6em]{0cm}{1.0em}\beta~[\%/mbar]$ & $-0.0555\pm$0.0010(stat.)$\pm$0.0009(sys.) &  &  &\\
			\cline{2-7}
			& \multirow{2}{*}{Middle} & $\rule{0cm}{1.2em}\alpha$ & $\phantom{-}0.0012\pm$0.0051(stat.)$\pm$0.0021(sys.) & \multirow{2}{*}{10.5$\pm$0.6}  & \multirow{2}{*}{11.0$\pm$0.7} & \multirow{2}{*}{0.951$\pm$0.002}\\
			& & $\rule[-0.6em]{0cm}{1.0em}\beta~[\%/mbar]$ & $-0.0559\pm$0.0015(stat.)$\pm$0.0001(sys.) &  & &\\
			\cline{2-7}	
			& \multirow{2}{*}{Down} & $\rule{0cm}{1.2em}\alpha$ & $-0.0114\pm$0.0030(stat.)$\pm$0.0015(sys.) & \multirow{2}{*}{10.5$\pm$0.6} & \multirow{2}{*}{11.1$\pm$0.7} & \multirow{2}{*}{0.952$\pm$0.002}\\
			& & $\rule[-0.6em]{0cm}{1.0em}\beta~[\%/mbar]$ & $-0.0588\pm$0.0009(stat.)$\pm$0.0000(sys.) &  & &\\
			\hline
			
			\multirow{7}{*}{Horizontal}
			& \multirow{2}{*}{Up} & $\rule{0cm}{1.1em}\alpha$ & $-0.2526\pm$0.0039(stat.)$\pm$0.0014(sys.) & \multirow{2}{*}{4.2$\pm$1.0} & \multirow{2}{*}{17.1$\pm$4.4} & \multirow{2}{*}{0.269$\pm$0.002}\\
			& & $\rule[-0.6em]{0cm}{1.0em}\beta~[\%/mbar]$ & $-0.0902\pm$0.0012(stat.)$\pm$0.0002(sys.) &  &  &\\
			\cline{2-7}
			& \multirow{2}{*}{Middle} & $\rule{0cm}{1.2em}\alpha$ & $-0.2623\pm$0.0060(stat.)$\pm$0.0019(sys.) & \multirow{2}{*}{4.2$\pm$1.0}  & \multirow{2}{*}{17.0$\pm$4.4} & \multirow{2}{*}{0.268$\pm$0.002}\\
			& & $\rule[-0.6em]{0cm}{1.0em}\beta~[\%/mbar]$ & $-0.0881\pm$0.0018(stat.)$\pm$0.0010(sys.) &  & &\\
			\cline{2-7}
			& \multirow{2}{*}{Down} & $\rule{0cm}{1.2em}\alpha$ & $-0.2734\pm$0.0035(stat.)$\pm$0.0017(sys.) & \multirow{2}{*}{4.2$\pm$1.0} & \multirow{2}{*}{16.9$\pm$4.5} & \multirow{2}{*}{0.265$\pm$0.002}\\
			& & $\rule[-0.6em]{0cm}{1.0em}\beta~[\%/mbar]$ & $-0.0939\pm$0.0010(stat.)$\pm$0.0002(sys.) &  & &\\
			\hline
		\end{tabular}
	\end{center}
	\end{table*}
	
\begin{acknowledgements}
	The DANSS collaboration expresses the deepest gratefulness to the staff and the top management of the Kalinin nuclear power plant for the continuous help and support during the experiment. The radiation safety department and the division of thermal measurements and automation deserve special thanks for the assistance with administrative procedures. This work would be totally impossible without the participation of the team from the laboratory of the reactor physics, who supplied the experiment with the most updated information on the reactor status and contributed to fruitful discussions.
	
	The design and construction of the DANSS setup became possible due to the support from the State Atomic Energy Corporation ’RosAtom’ in the framework of the state contracts No H.4x.44.90.13.1119 and No H.4x.44.9B.16.1006. The long-term operation of the detector, the experimental data taking and their analysis are carried out under the grants No 17-12-01145 and No 17-12-01145n of the Russian Scientific Foundation.
\end{acknowledgements}



\end{document}